\documentclass[aps,prb,twocolumn,footinbib,groupedaddress,longbibliography]{revtex4-2}
\usepackage{graphicx}
\usepackage{subfigure}
\usepackage{dcolumn}
\usepackage{bm}
\usepackage{times}
\usepackage{amsmath}
\usepackage{amssymb}
\usepackage{float}
\usepackage{color}
\usepackage{multirow}
\usepackage{url,hyperref}
\usepackage[normalem]{ulem}

\usepackage{graphicx}
\usepackage{epstopdf}
\usepackage{float}
\usepackage{amsmath}
\usepackage{upgreek}

\begin{document}
\title{Enhancing far-field thermal radiation by Floquet engineering}
\author{Huimin Zhu,$^{1,2}$ Yuhua Ren,$^{2}$ Hui Pan,$^{2}$ Gaomin Tang,$^{3}$ Lei Zhang,$^{1,4}$ and Jian-Sheng Wang$^{2,*}$}
\affiliation{$^1$State Key Laboratory of Quantum Optics Technologies and Devices, Institute of Laser Spectroscopy, Shanxi University, Taiyuan 030006, China\\
  $^2$Department of Physics, National University of Singapore, Singapore 117551, Republic of Singapore\\
  $^3$Graduate School of China Academy of Engineering Physics, Beijing 100193, China\\
  $^4$Collaborative Innovation Center of Extreme Optics, Shanxi University, Taiyuan 030006, China}

\bigskip

\begin{abstract}
Time modulation introduces a dynamic degree of freedom for tailoring thermal radiation beyond the limits of static materials. Here we investigate far-field thermal radiation from a periodically time-modulated SiC film under the Floquet nonequilibrium Green's function framework. We show that time modulation enables radiative energy transfer into the far field that surpasses the limit imposed by the equilibrium thermal fluctuations. This enhancement originates from the modulation-induced coupling between evanescent surface phonon polaritons and propagating modes, effectively bridging the energy and momentum mismatch through frequency conversion. Notably, even at zero temperature, the film emits a finite radiative heat flux due to nonequilibrium photon occupation generated by the modulation. The radiative output grows with increasing modulation strength, highlighting the role of external work in driving far-field emission. These results establish time modulation as an effective mechanism for bridging near-field and far-field regimes, opening new pathways for active thermal radiation control.
\end{abstract}

\maketitle

\textit{Introduction.}-- Radiative heat transfer exhibits strong scale dependence, with its behavior changing markedly with the distance between objects~\cite{chapuis2023thermal}. Consequently, controlling thermal radiation across different length scales is critical for a wide range of applications~\cite{baranov2019nanophotonic,inoue2018spectral,elzouka2018meshed,latella2021smart,song2015near}. In the far-field regime, where the separation $d$ exceeds the thermal wavelength $\lambda_{\text{th}}$, radiative heat transfer is well described by Planck's law~\cite{Planck1901} and the Stefan-Boltzmann law~\cite{bergman2011fundamentals}, with energy exchange dominated by propagating electromagnetic waves~\cite{Biehs2021Near}. In contrast, in the near-field regime ($d \ll \lambda_{\text{th}}$), evanescent modes enable photon tunneling~\cite{Rousseau2009Radiative,lim2018tailoring,shen2009surface,Zhao2017near}, leading to radiative heat fluxes that can exceed the blackbody limit by several orders of magnitude~\cite{Zhang_book}. This enhancement, driven by surface electromagnetic modes, underpins key applications including thermophotovoltaics~\cite{messina2013graphene,PARK2008305}, thermal rectification~\cite{Linxiao2013}, noncontact refrigeration~\cite{Kaifeng2016}, and thermal transistors~\cite{Abdallah2014}. However, the rapid spatial decay of surface waves away from interfaces~\cite{cravalho1967effect} restricts near-field effects to sub-wavelength distances, severely limiting their practical reach. Bridging near- and far-field regimes—thus enabling the exploitation of near-field phenomena over extended distances—remains a central challenge~\cite{Gallinet11,Odebowale2421051}. Various strategies have been explored, including nano-coatings~\cite{Tachikawa2024}, grating structures~\cite{greffet2002coherent}, active gain media~\cite{ding2016active1}, optical antennas~\cite{schuller2009optical}, and thermal extraction lenses~\cite{yu2013enhancing}, to facilitate the coupling of near-field modes into the far field.

Recently, time-varying metamaterials have attracted growing attention for enabling unprecedented light-matter interactions by exploiting temporal modulation~\cite{taravati2021pure,shcherbakov2019photon,ramaccia2019phase,zhou2020broadband,pang2021adiabatic,sounas2017non,cardin2020surface,barati2022optical,galiffi2020wood,tsai2022surface}. While time-modulation approaches have been proposed to tailor thermal radiation~\cite{tang2024modulating,coppens2017spatial,buddhiraju2020photonic,yu2024time,ordonez2024far,vazquez2023incandescent,vertiz2025dispersion,liberal2024can}, existing studies primarily address near-field or far-field thermal radiation separately. The potential of time modulation to control the coupling between near-field and far-field regimes remains largely unexplored. This raises a natural and compelling question: Could time modulation serve as an additional mechanism to harness near-field effects for far-field thermal radiation? If so, can far-field emission be significantly enhanced via near-field energy transfer?

In this Letter, we investigate far-field thermal radiation from a system subject to periodic time modulation, employing the nonequilibrium Green's function formalism. As a representative example, we consider a periodically modulated SiC film and demonstrate that time modulation can significantly enhance far-field thermal emission beyond the level set by intrinsic fluctuating currents. This enhancement arises from the coupling of evanescent modes to propagating modes, with the Floquet drive effectively shifting the near-field evanescent modes into the far-field propagating modes. We further examine the temperature dependence of the energy currents and reveal the critical role of the work performed by the modulation in sustaining far-field emission. Notably, time modulation can induce thermal radiation even from a zero-temperature object, driving it out of equilibrium and enabling emission into the vacuum. Finally, we show that the far-field thermal radiation can be further amplified by increasing the modulation strength.

\begin{figure}
\includegraphics[width=\columnwidth]{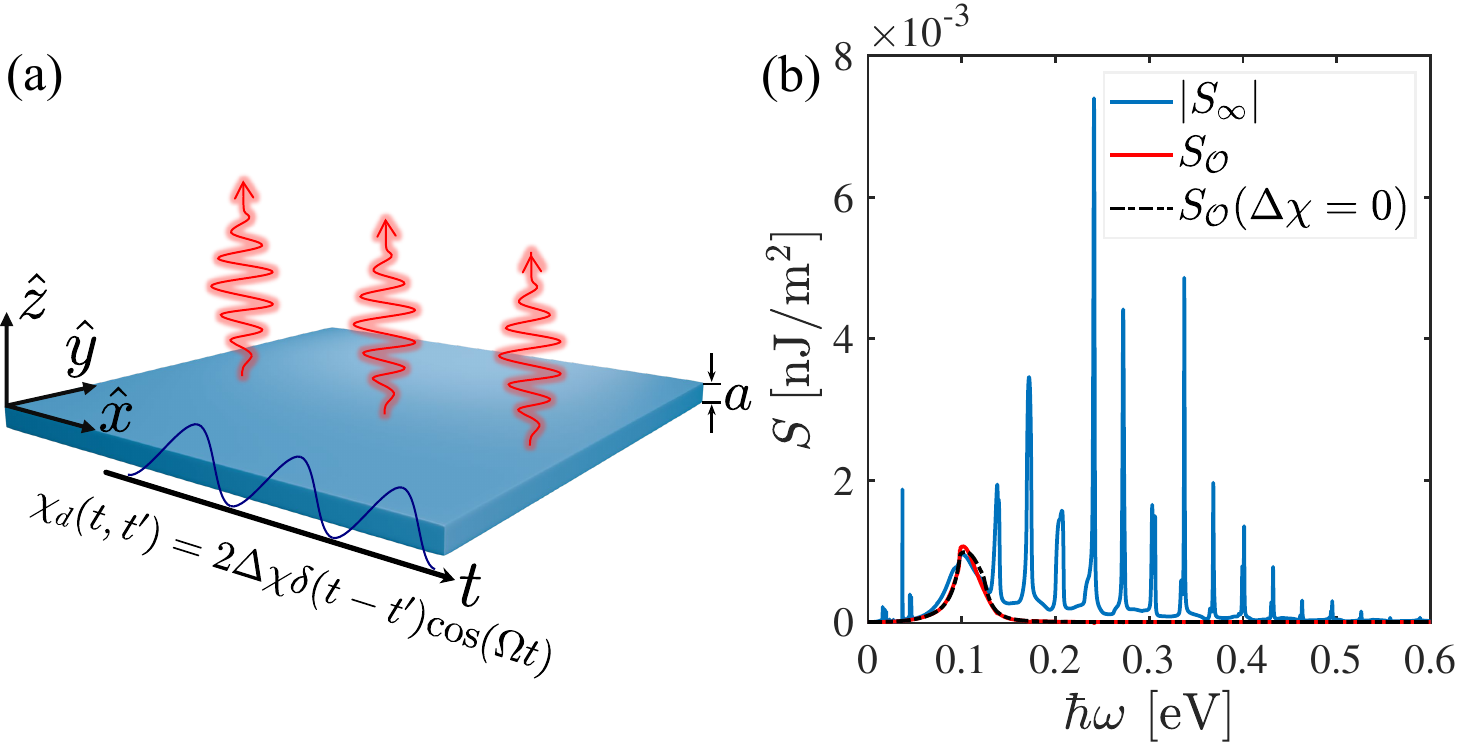} \\
\caption{(a) Schematic of thermal radiation from a film with thickness $a$, temperature $T$, and electric susceptibility periodically modulated by $\chi_d(t,t')$. (b) Thermal emission spectra $|S_{\infty}|$ and $S_\mathcal{O}$ as a function of $\hbar \omega$ for $\Delta \chi = 2.5$, $\hbar \Omega = 30.9\,\text{meV}$, and $T = 300\,\text{K}$. For comparison, the thermal emission spectrum $S_\mathcal{O}(\Delta \chi = 0)$ in the absence of modulation, obtained from ${\text{Eq.\,(8)}}$, is also shown.}
\label{fig1}
\end{figure}

\textit{System.}-- As illustrated in Fig.~\ref{fig1}(a), we consider a time-modulated film located in the $xy$ plane and maintained at temperature $T$. The surrounding environment is modeled as thermal baths at infinity and zero temperature. Under time modulation, the electric susceptibility can be decomposed as $\chi(t,t')= \chi_0(t - t') + \chi_d(t,t')$. Here, $\chi_0(t - t')$ represents the time-invariant component, associated with the intrinsic permittivity $\epsilon(t-t')$ of the film via the relation $ \chi_0(t-t') =\epsilon(t-t') - 1 $. The time-modulated component~\cite{vazquez2023incandescent}  $\chi_d(t,t') = 2\Delta\chi\, \delta(t - t') \cos(\Omega t)$ is characterized by modulation strength $\Delta\chi$ and frequency $\Omega$. Since the driven part $\chi_d(t,t')$ is dissipationless, all thermal radiation originates from the equilibrium component $\chi_0(t-t')$~\cite{yu2023manipulating}.

The far-field thermal radiation of the time-modulated film originates from two sources: the thermal radiation emitted by the fluctuating current sources contributed by the equilibrium component $\chi_0(t - t')$ of the susceptibility, and the energy injected into the emitted field by the work done on the electromagnetic field through time modulation $\chi_d(t,t')$. Their combined contribution to the far-field thermal radiation can be calculated by evaluating the energy flux into the thermal baths at infinity. Therefore, for the system shown in Fig.~\ref{fig1}(a), energy conservation must be satisfied~\cite{wang2023transport}:

\begin{align}
 {I}_{\infty}+{I}_{\mathcal{O}} + {I}_d =0,
\end{align}
where the subscripts $\infty$, $\mathcal{O}$, and $d$ refer to the baths at infinity, the equilibrium contribution of the film, and the portion driven by time modulation, respectively. Accordingly, $I_\infty = I_{+\infty} + I_{-\infty}$ is the total energy emitted by the two baths at $+\infty$ and $-\infty$, with ${I}_{+\infty}={I}_{-\infty}$ due to the  spatial symmetry of the film in the $z$-direction. The term  ${I}_{\mathcal{O}}$ is the radiation emitted by the fluctuating current sources in the film, and ${I}_d$ is the energy injected into the emitted field through time modulation. 

\begin{figure*}
\includegraphics[scale=0.4]{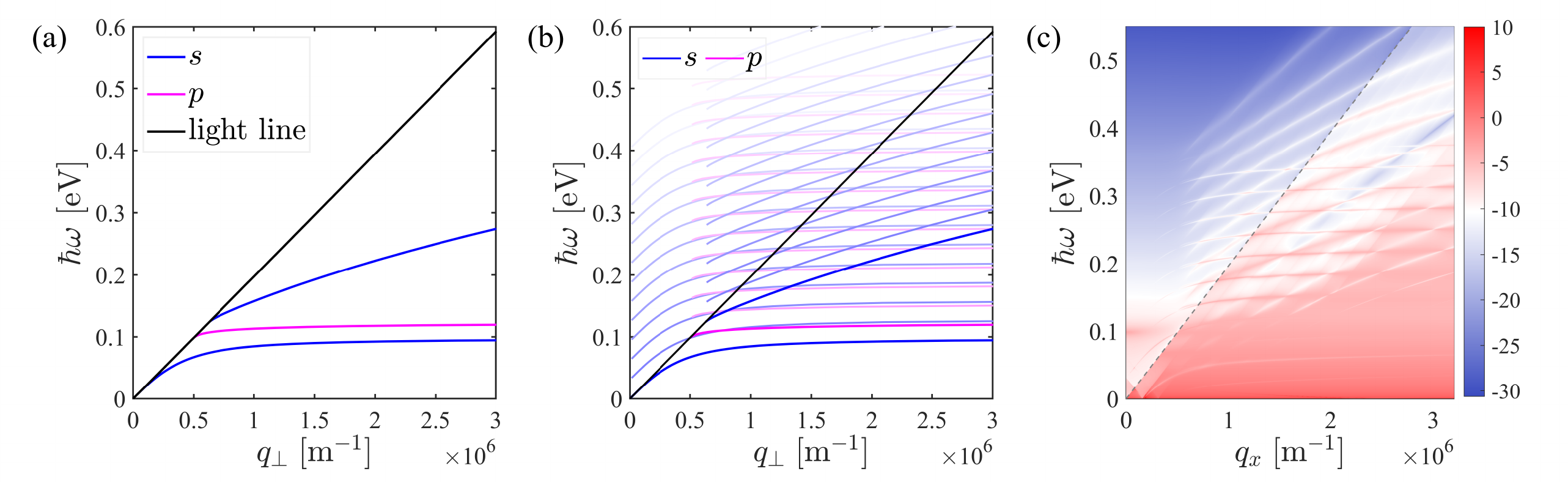} \\
\caption{(a) Dispersion relations of $s$- and $p$-polarized surface phonon polaritons supported by the SiC film at $\Delta\chi = 0$. (b) Schematic of the coupling between surface modes and propagating modes under time modulation for $n \geq 0$. Curves with larger values of $n$ are depicted in lighter colors. (c) Effective photon distribution, $\log_{10} \bar{N}$, as a function of $q_x$ and $\hbar \omega$ at $q_y = 0$. The parameters used in (b) and (c) are $\Delta\chi = 2.5$, $\hbar \Omega = 30.9\,\text{meV}$, and $T = 300\,\text{K}$.}
\label{fig2}
\end{figure*}

Using Poynting's theorem~\cite{griffiths2023} together with the nonequilibrium Green's function formalism~\cite{wang2008quantum,zhang2022microscopic,wang2014nonequilibrium}, the energy currents per unit area ${I}_{\infty}$ and ${I}_{\mathcal{O}}$ can be expressed as (with detailed derivations provided in the Supplemental Materials~\cite{supple})
\begin{align}
 {I} = \int_0^{\infty} \frac{d\omega}{2\pi} S(\omega),  \label{I}
\end{align}
with $I \in \{I_{\infty}, I_{\mathcal{O}}\}$ and the corresponding spectral functions $S \in \{S_{\infty}, S_{\mathcal{O}}\}$. The spectral functions are given by:
\begin{align}
& {S}_\infty = 2\hbar \omega \sum \limits_{n} \int_{|\mathbf{q}_\perp| < k_0} \frac{d^2 \mathbf{q}_\perp}{(2\pi)^2}  \bigg[\frac{ {\rm Im} [\Pi_0^r(\omega_n)]}{\epsilon_0 c^2 k}  N(T_{\mathcal{O}},\omega_n) \notag \\
& \times \left( |t^s_n|^2 + \frac{k^2 }{k_0^2} |t^p_n|^2 \right) - \left( \frac{k_n} {k}|t^s_n|^2 +\frac{k \omega_n^2}{k_n \omega^2}|t^p_n|^2 \right) \theta(-\omega_n) \bigg] , \label{Iinfty} 
\end{align}
\begin{align}
&{S}_{\mathcal{O}} = \frac{\hbar \omega}{\epsilon_0 c^2} \sum\limits_{n} \int_{|\mathbf{q}_\perp|<k_0} \frac{d^2 \mathbf{q}_\perp}{(2\pi)^2} {\rm Im}[\Pi_0^r(\omega)] \Bigg\{
\frac{{\rm Im}[\Pi_0^r (\omega_n)]}{\epsilon_0 c^2 k^2} \notag \\
& \quad \times \left(|t^s_n|^2 + \frac{k^4}{k_0^4}|t^p_n|^2 \right) \Big[ N(T_{\mathcal{O}}, \omega)-N(T_{\mathcal{O}}, \omega_n) \Big]  \notag \\
& \quad - \left( \frac{2k_n}{k^2} |t^s_n|^2+ \frac{2\omega_n^2 k^2} {\omega^2 k_0^2 k_n} |t^p_n|^2 \right) \Big[ N(T_{\mathcal{O}}, \omega) + \theta(-\omega_n) \Big] \Bigg\}. \label{ISiC}
\end{align}
The in-plane wavevector is denoted by ${\mathbf{q}_\perp} = (q_x, q_y)$, and $\omega_n = \omega + n \Omega$ with $n \in \mathbb{Z}$. The magnitude of the out-of-plane wavevector in vacuum is $k = \sqrt{k_0^2 - q_\perp^2}$, where $k_0 = \omega/c$ and $q_\perp = |{\mathbf{q}_\perp}|$. Similarly, $k_n = \sqrt {(\omega_n/c)^2 - q_\perp^2}$. The Bose-Einstein distribution function is $N(T, \omega) = \left[\exp(\hbar \omega / k_B T) - 1 \right]^{-1}$,  the Heaviside step function is denoted by $\theta(\omega)$, and $T=T_{\mathcal{O}}$ is the temperature of the film.  The polarization function is given by:
\begin{align}
\Pi_0^r (\omega_n) = -a \epsilon_0 \omega_n^2 \big[ \epsilon (\omega_n) - 1 \big],
\end{align}
where $\epsilon_0$ is the vacuum permittivity. The transmission coefficients for $s$- and $p$-polarization in Floquet space are given by:
\begin{align}
    \bm{t}^s &= \big[ \mathbf{I} - {i a}/{(2 \hbar^2 c^2)} \bm{\mathcal{E}} \bm{\chi} \bm{\mathcal{E}} \bm{k}^{-1} \big]^{-1}, \\
    \bm{t}^p &= \big[ \mathbf{I} - {i a} \bm{\mathcal{E}} \bm{\chi} \bm{\mathcal{E}}^{-1} \bm{k} /2 \big]^{-1}.
\end{align}
Here, $\bm{t}^{s,p}$ is a matrix of size $l \times l$ (with $l$ odd), where $t^{s,p}_n = \bm{t}^{s,p}[(l+1)/2, (l+1)/2 + n]$ with $(l+1)/2 > |n|$. The identity matrix is denoted by $\mathbf{I}$, $\bm{\chi}$ is the Floquet representation of $\chi(t,t')$, and the entries of the matrix $\bm{\mathcal{E}}$ are $\mathcal{E}_n \delta_{mn}$ with $\mathcal{E}_n = \hbar \omega_n$. The diagonal matrix $\bm{k}$ has elements $k_{mn} = \delta_{mn} k_n$, where $k_n = \sqrt{\left(\omega_n/c\right)^2 - q_\perp^2}$ if $|\omega_n| < c q_\perp$, and $k_n = \text{sgn}(\omega_n)\sqrt{\left(\omega_n/c\right)^2 - q_\perp^2} $ if $|\omega_n| > c q_\perp$. The far-field and near-field regimes correspond to $q_\perp < k_0$ and $q_\perp > k_0$, respectively. 
The energy currents $I_{\infty} < 0$ and $I_{\mathcal{O}} > 0$, with the sign convention that emitted energy is positive, and absorbed energy is negative. The expression for the modulation-induced energy current $I_d$ is provided in the Supplemental Material~\cite{supple}.
In the absence of time modulation, $I_d$ vanishes, and both $I_{\mathcal{O}}$ and $-I_\infty$ reduce to the expression describing thermal  emission~\cite{Matthias2012} with
\begin{align}\label{I_undrive}
{I} = &\int_{0}^{\infty} \frac{d\omega}{2\pi} \int_{|\mathbf{q}_\perp|<k_0} \frac{d^2 \mathbf{q}_\perp}{(2\pi)^2}
  \sum\limits_{j = s,p} 2\hbar\omega {N}(T,\omega) \notag\\
  & \times \left( 1 -|r^j|^2 - |t^j|^2 \right),
\end{align}
where $r^j$ and $t^j$ are the reflection and transmission coefficients for the undriven film, respectively, and they satisfy $1+r^j=t^j$.

\textit{Numerical results.}-- We consider a SiC film with thickness $a = 0.529\, \mathrm{\upmu}{\mathrm m}$. Its permittivity is described by a Drude-Lorentz model~\cite{vazquez2023incandescent,joulain2005surface}: $\varepsilon(\omega) = \varepsilon_{\infty} (\omega_{L}^{2} - \omega^{2} - i\Gamma \omega)/(\omega_{T}^{2} - \omega^{2} - i\Gamma\omega)$, where $\varepsilon_{\infty}= 6.7$ is the high-frequency dielectric constant, $\omega_{L} / 2\pi  = 29.1 \,\mathrm{THz}$ and $\omega_{T} / 2\pi = 23.8 \, \mathrm{THz}$ are the longitudinal and transverse optical phonon frequencies, and $\Gamma / 2\pi = 0.14 \, \mathrm{THz}$ is the damping rate.
In Fig.~\ref{fig1}(b), we present the far-field thermal emission spectrum $|S_\infty|$ of the SiC film and the thermal emission spectrum $S_{\mathcal{O}}$ due to the equilibrium part, both under time modulation with $\Delta\chi = 2.5$, $\hbar\Omega = 30.9\, \text{meV}$, and $T = 300\, \text{K}$. For comparison, we also show the thermal emission spectrum $S_{\mathcal{O}} (\Delta\chi = 0)$ in the absence of modulation, calculated from Eq.~\eqref{I_undrive} and arising from equilibrium thermal fluctuations of current sources within the SiC film. Under time modulation, the far-field thermal emission from the SiC film is significantly enhanced compared to the radiation emitted by the fluctuating current sources within the material that are driven out of equilibrium, especially in the high-frequency regime. Moreover, the frequency spacing between adjacent peaks of the $|S_\infty|$ spectrum precisely corresponds to the modulation frequency. Notably, both the unmodulated spectrum $S_\mathcal{O}(\Delta \chi = 0)$ and the modulated spectrum $S_\mathcal{O}$ at $\Delta \chi = 2.5$ due to the equilibrium part exhibit a single dominant peak with highly similar spectral profiles, while showing discernible differences. This result indicates that the nonequilibrium state induced by periodic driving has a weak but observable effect on the equilibrium thermal radiation of the material.

To further elucidate the underlying physics of the modulation-induced spectral enhancement, we examine the surface phonon polaritons (SPhPs) supported by the SiC film. The dispersion relations for the transverse electric ($s$-polarized) and transverse magnetic ($p$-polarized) modes are given by~\cite{mikhailov2007new}:
\begin{align}
q_\perp^2 =& k_0^2 + \frac{1}{4} a^2 \big[ \epsilon(\omega)-1 \big]^2 k_0^4,\label{TE} \\
q_\perp^2 =& k_0^2 + \frac{4}{a^2 \big[ \epsilon(\omega)-1 \big]^2}.\label{TM}
\end{align}
It is important to note that, for the stable existence of modes, the dispersion relations must satisfy the condition $ik < 0$, where $k=\sqrt{k_0^2-q_{\perp}^2}$. Additional details are provided in the Supplemental Material~\cite{supple}.

Figure~\ref{fig2}(a) shows the dispersion relations for the surface modes on an unmodulated SiC film, along with the vacuum light line $\omega = c k_0$ (black line). The surface modes, due to their higher energy density, significantly enhance energy transfer in the near field~\cite{polder1971theory}. However, these modes are inaccessible in the far field, which is the focus of our study. From the dispersion relations, the in-plane momentum $q_\perp$ of the surface modes exceeds the momentum $k_0$ of photons propagating in free space, whereas for the propagating mode, $q_\perp < k_0$. This suggests that coupling between the two types of modes requires bridging the momentum or energy difference. Under time modulation, the time reversal symmetry is broken, resulting in $\omega' = \omega_{s,p}(q_\perp) + n\Omega$, where $n$ is an integer. Here, $\omega_{s,p}(q_\perp)$ denotes the frequency associated with the wavevector $q_\perp$ on the dispersion curves, with the subscripts $s$ and $p$ corresponding to the $s$-polarized and $p$-polarized modes, respectively. Figure~\ref{fig2}(b) illustrates the coupling of the surface modes to the propagating regime under time modulation at $\Delta\chi = 2.5$ and $\hbar \Omega = 30.9\, \text{meV}$. This figure provides a schematic representation rather than the actual dispersion curves under time modulation. For a given wave vector $q_\perp$, the frequency of the propagating mode exceeds that of the surface mode, enabling access to the surface mode in the far-field regime when $\omega' > \omega_{s,p}$. Therefore, we focus on the dispersion curves for $n \geq 0$, where the $n = 0$ dispersion curves correspond to the unmodulated case shown in Fig.~\ref{fig2}(a).

\begin{figure*}
\centering
\includegraphics[scale=0.4]{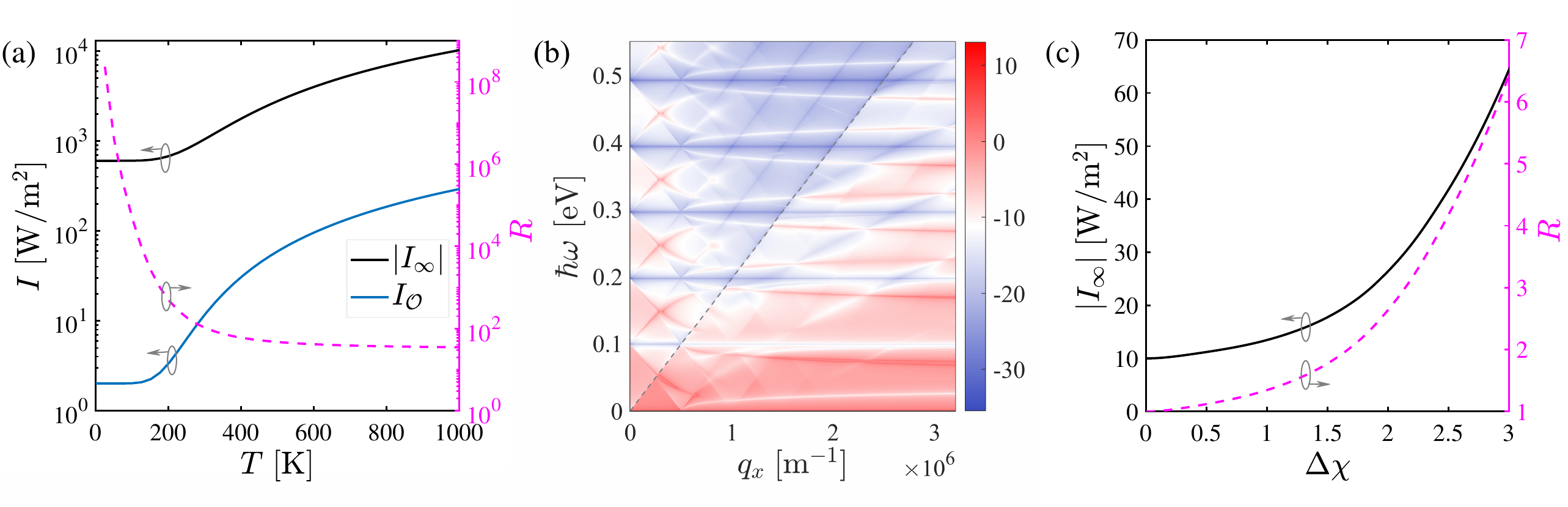} \\  
\caption{(a) The energy currents $|{I}_{\infty}|$ and ${I}_{\mathcal{O}}$ as functions of temperature $T$ at $\Delta\chi=2.5$ and $\hbar\Omega=\hbar\omega_T$ ($98.4\, \text{meV}$, left axis). Here, ${I}_{\infty}$ (black line) represents the energy emitted by the baths at infinity and ${I}_{\mathcal{O}}$ (blue line) denotes the thermal emission from the equilibrium part of the SiC film. The enhancement factor $R=|{I}_{\infty}(T,\Delta\chi=2.5)/{I}_{\infty}(T,\Delta\chi=0)|$ versus temperature $T$ is also shown (right axis). (b) The effective photon distribution $\log_{10}\bar{N}$ plotted against $q_x$ and $\hbar\omega$ at $q_y=0$, $T = 0$, $\Delta\chi=2.5$, and  $\hbar \Omega=\hbar \omega_T$. The gray dotted line represents the vacuum light line. (c) The energy current $|I_{\infty}|$ as a function of modulation strength $\Delta\chi$ at $\hbar\Omega=30.9\, \text{meV}$ and $T = 300\,\text{K}$, shown on the left axis, with the corresponding enhancement factor $R = |I_{\infty}(T=300\,\text{K}, \Delta\chi) \big/ I_{\infty}(T=300\,\text{K}, \Delta\chi=0)|$ displayed on the right axis.}
\label{fig3}
\end{figure*}

To illustrate the coupling mechanism described above, we analyze the effective photon distribution of the SiC film, defined as~\cite{Pan2025Asymmetry}:
\begin{align} \label{N}
\bar{N} ({\mathbf{q}_\perp}, \omega) = \frac{ {\rm Tr} \big\{ {\rm Im} [\bm{D}^<_{00} ({\mathbf{q}_\perp}, \omega)] \big\}}{2\,{\rm Tr} \big\{ {\rm Im} [\bm{D}^r_{00} ({\mathbf{q}_\perp}, \omega)] \big\}},
\end{align}
where the trace (sum over the directions) is taken from the ``00'' block of the Floquet matrix. The lesser Green's function is given by $\bm{D}^< = \bm{D}^r (\bm{\Pi}^< + \bm{\Pi}^<_\infty) \bm{D}^a$ in Floquet space. Here, $\bm{D}^r$ and $\bm{D}^a$ are the retarded and advanced Green's functions, respectively, and $\bm{\Pi}^<$ denotes the self-energy of the SiC film, while $\bm{\Pi}^<_\infty$ represents the energy dissipated into the vacuum, given by:
\begin{align}
\bm{\Pi}^<_\infty = 2 i \epsilon_0 c^2 \bm{k} \left[ {\mathbf I}_\perp + \frac{1}{ \bm{k}^2} \left( {\mathbf q}_\perp {\mathbf q}^{T}_\perp \right) \right] \theta(-\omega_n).
\end{align}
In the unmodulated case, Eq.~\eqref{N} reduces to the standard Bose-Einstein distribution in thermal equilibrium. In the modulated case, we present the effective photon distribution $\log_{10} \bar{N}$ in Fig.~\ref{fig2}(c). The far-field regime within the light cone exhibits multiple bright bands. As discussed in Fig.~\ref{fig2}(b), these bright bands arise from the coupling of surface modes to the propagating regime via frequency modulation, significantly contributing to the thermal radiation. This mechanism accounts for the multi-peak structure observed in the far-field thermal emission spectrum $|{S}_\infty|$ shown in Fig.~\ref{fig1}(b).

We next examine the temperature dependence of the photon energy currents. Using Eqs.~\eqref{I}-\eqref{ISiC}, we show ${I}_{\mathcal{O}}$ and the absolute value of ${I}_{\infty}$ as functions of the SiC film temperature $T$ at $\Delta\chi=2.5$ and $\hbar \Omega=\hbar \omega_T$ ($98.4\, \text{meV}$) on the left axis of Fig.~\ref{fig3}(a), respectively. Strikingly, the far-field thermal radiation $|I_{\infty}|$ exceeds the equilibrium contribution $I_{\mathcal{O}}$ across the entire temperature range considered. To quantify the impact of time modulation, we define an enhancement factor $R=|{I}_{\infty}(T,\Delta\chi=2.5)/{I}_{\infty}(T,\Delta\chi=0)|$,  shown on the right axis of Fig.~\ref{fig3}(a) as a function of temperature. The results reveal a pronounced enhancement of far-field emission induced by time modulation, particularly at low temperatures. Although the enhancement diminishes with increasing temperature, it remains substantial, sustaining an order-of-magnitude increase even at elevated temperatures.

Remarkably, ${I}_{\mathcal{O}}$ approaches a constant rather than zero at $T=0$, which implies that the SiC film still radiates finite energy to the baths at infinity due to time modulation. This seems to contradict the third law of thermodynamics, which states that a system should be in its ground state at $T=0$, where no spontaneous energy radiation should occur. This is indeed the case for the unmodulated case, where the photon occupation number $N(T=0,\omega)=0$ with $\omega>0$. However, for the modulated case, $\omega$ is replaced by the modulation frequency $\omega_n$, and $N(T=0,\omega_n)=-1$ for $\omega_n<0$. This is a crucial distinction between the modulated and unmodulated cases.
To understand this effect more intuitively, we present the effective photon distribution under periodic modulation at $T=0$, $\Delta\chi=2.5$, and $\hbar\Omega=\hbar\omega_T$ in Fig.~\ref{fig3}(b). The results show that time modulation significantly affects the photon distribution of the SiC film, driving it into a nonequilibrium state, while the baths at infinity remain in equilibrium. Therefore, the energy radiation at $T=0$ is a result of the SiC film in a nonequilibrium state. Furthermore, for the entire time-modulated system at $T=0$, time modulation drives the SiC film into a nonequilibrium state while also performing work on the electromagnetic radiation emitted by the film. Ultimately, these two effects contribute $|{I}_{\infty}|=602.9\,$W$/$m$^{2}$ of energy to be absorbed by the baths. This energy is non-negligible and comparable to the blackbody radiation at $T=321\,\text{K}$.

Finally, we investigate the effect of the modulation strength on the far-field thermal radiation at $\hbar\Omega=30.9\, \text{meV}$ and $T = 300\,\text{K}$, as shown on the left axis of Fig.~\ref{fig3}(c). The modulation strength determines the intensity of the work performed by the time modulation on the system. As $\Delta\chi$ increases, the  electromagnetic field emitted responds more strongly to the time modulation, resulting in increased work performed, which enhances the far-field thermal radiation. This enhancement is quantified on the right axis of Fig.~\ref{fig3}(c) through the enhancement factor $R = |I_{\infty}(T=300\,\text{K}, \Delta\chi) \big/ I_{\infty}(T=300\,\text{K}, \Delta\chi=0)|$. The enhancement factor increases with increasing $\Delta \chi$, reaching up to about a $6.4$-fold enhancement at $\Delta \chi=3$, demonstrating the ability of modulation strength to enhance far-field thermal radiation.  

\textit{Conclusion.}-- In summary, we have studied the far-field thermal radiation of a SiC film under periodic modulation and obtained the expressions describing the far-field thermal radiation. Our results show that time modulation significantly enhances far-field thermal radiation. This arises from the coupling between surface modes and propagating modes, and is primarily contributed by the work done through time modulation. We have also quantitatively demonstrated the significant enhancement of far-field thermal radiation due to time modulation through the enhancement factor and shown that the far-field thermal radiation can be enhanced by increasing the modulation strength. Our results suggest promising opportunities for harnessing time modulation to actively control thermal emission at the nanoscale and dynamically regulate energy transfer in the far-field regime. 

\begin{acknowledgments}
H. P. and J.-S. W. acknowledge support from the Ministry of Education, Singapore, under the Academic Research Fund (FY2022). H. M. Z. and L. Z. gratefully acknowledge the support from the National Natural Science Foundation of China (Grant No. 12474047, 12174231), the Fund for Shanxi ``1331 Project", Research Project Supported by Shanxi Scholarship Council of China. G.T. is supported by National Natural Science Foundation of China (Grants No. 12088101 and No. 12374048) and NSAF (Grant No. U2330401).
\end{acknowledgments}

\bigskip

\bibliography{Farfield_Floquet}{}

\begin{thebibliography}{63}%
\makeatletter
\providecommand \@ifxundefined [1]{%
 \@ifx{#1\undefined}
}%
\providecommand \@ifnum [1]{%
 \ifnum #1\expandafter \@firstoftwo
 \else \expandafter \@secondoftwo
 \fi
}%
\providecommand \@ifx [1]{%
 \ifx #1\expandafter \@firstoftwo
 \else \expandafter \@secondoftwo
 \fi
}%
\providecommand \natexlab [1]{#1}%
\providecommand \enquote  [1]{``#1''}%
\providecommand \bibnamefont  [1]{#1}%
\providecommand \bibfnamefont [1]{#1}%
\providecommand \citenamefont [1]{#1}%
\providecommand \href@noop [0]{\@secondoftwo}%
\providecommand \href [0]{\begingroup \@sanitize@url \@href}%
\providecommand \@href[1]{\@@startlink{#1}\@@href}%
\providecommand \@@href[1]{\endgroup#1\@@endlink}%
\providecommand \@sanitize@url [0]{\catcode `\\12\catcode `\$12\catcode
  `\&12\catcode `\#12\catcode `\^12\catcode `\_12\catcode `\%12\relax}%
\providecommand \@@startlink[1]{}%
\providecommand \@@endlink[0]{}%
\providecommand \url  [0]{\begingroup\@sanitize@url \@url }%
\providecommand \@url [1]{\endgroup\@href {#1}{\urlprefix }}%
\providecommand \urlprefix  [0]{URL }%
\providecommand \Eprint [0]{\href }%
\providecommand \doibase [0]{http://dx.doi.org/}%
\providecommand \selectlanguage [0]{\@gobble}%
\providecommand \bibinfo  [0]{\@secondoftwo}%
\providecommand \bibfield  [0]{\@secondoftwo}%
\providecommand \translation [1]{[#1]}%
\providecommand \BibitemOpen [0]{}%
\providecommand \bibitemStop [0]{}%
\providecommand \bibitemNoStop [0]{.\EOS\space}%
\providecommand \EOS [0]{\spacefactor3000\relax}%
\providecommand \BibitemShut  [1]{\csname bibitem#1\endcsname}%
\let\auto@bib@innerbib\@empty
\bibitem [{\citenamefont {Chapuis}\ \emph {et~al.}(2023)\citenamefont
  {Chapuis}, \citenamefont {Lee},\ and\ \citenamefont
  {Rodriguez}}]{chapuis2023thermal}%
  \BibitemOpen
  \bibfield  {author} {\bibinfo {author} {\bibfnamefont {Pierre-Olivier}\
  \bibnamefont {Chapuis}}, \bibinfo {author} {\bibfnamefont {Bong~Jae}\
  \bibnamefont {Lee}}, \ and\ \bibinfo {author} {\bibfnamefont {Alejandro}\
  \bibnamefont {Rodriguez}},\ }\bibfield  {title} {\enquote {\bibinfo {title}
  {Thermal radiation at the nanoscale and applications},}\ }\href {\doibase
  10.1063/5.0186406} {\bibfield  {journal} {\bibinfo  {journal} {Appl. Phys.
  Lett.}\ }\textbf {\bibinfo {volume} {123}},\ \bibinfo {pages} {220401}
  (\bibinfo {year} {2023})}\BibitemShut {NoStop}%
\bibitem [{\citenamefont {Baranov}\ \emph {et~al.}(2019)\citenamefont
  {Baranov}, \citenamefont {Xiao}, \citenamefont {Nechepurenko}, \citenamefont
  {Krasnok}, \citenamefont {Al{\`u}},\ and\ \citenamefont
  {Kats}}]{baranov2019nanophotonic}%
  \BibitemOpen
  \bibfield  {author} {\bibinfo {author} {\bibfnamefont {Denis~G}\ \bibnamefont
  {Baranov}}, \bibinfo {author} {\bibfnamefont {Yuzhe}\ \bibnamefont {Xiao}},
  \bibinfo {author} {\bibfnamefont {Igor~A}\ \bibnamefont {Nechepurenko}},
  \bibinfo {author} {\bibfnamefont {Alex}\ \bibnamefont {Krasnok}}, \bibinfo
  {author} {\bibfnamefont {Andrea}\ \bibnamefont {Al{\`u}}}, \ and\ \bibinfo
  {author} {\bibfnamefont {Mikhail~A}\ \bibnamefont {Kats}},\ }\bibfield
  {title} {\enquote {\bibinfo {title} {Nanophotonic engineering of far-field
  thermal emitters},}\ }\href {\doibase 10.1038/s41563-019-0363-y} {\bibfield
  {journal} {\bibinfo  {journal} {Nat. Mater.}\ }\textbf {\bibinfo {volume}
  {18}},\ \bibinfo {pages} {920--930} (\bibinfo {year} {2019})}\BibitemShut
  {NoStop}%
\bibitem [{\citenamefont {Inoue}\ \emph {et~al.}(2018)\citenamefont {Inoue},
  \citenamefont {Asano},\ and\ \citenamefont {Noda}}]{inoue2018spectral}%
  \BibitemOpen
  \bibfield  {author} {\bibinfo {author} {\bibfnamefont {Takuya}\ \bibnamefont
  {Inoue}}, \bibinfo {author} {\bibfnamefont {Takashi}\ \bibnamefont {Asano}},
  \ and\ \bibinfo {author} {\bibfnamefont {Susumu}\ \bibnamefont {Noda}},\
  }\bibfield  {title} {\enquote {\bibinfo {title} {Spectral control of
  near-field thermal radiation via photonic band engineering of two-dimensional
  photonic crystal slabs},}\ }\href {\doibase 10.1364/OE.26.032074} {\bibfield
  {journal} {\bibinfo  {journal} {Opt. Express}\ }\textbf {\bibinfo {volume}
  {26}},\ \bibinfo {pages} {32074--32082} (\bibinfo {year} {2018})}\BibitemShut
  {NoStop}%
\bibitem [{\citenamefont {Elzouka}\ and\ \citenamefont
  {Ndao}(2018)}]{elzouka2018meshed}%
  \BibitemOpen
  \bibfield  {author} {\bibinfo {author} {\bibfnamefont {Mahmoud}\ \bibnamefont
  {Elzouka}}\ and\ \bibinfo {author} {\bibfnamefont {Sidy}\ \bibnamefont
  {Ndao}},\ }\bibfield  {title} {\enquote {\bibinfo {title} {Meshed doped
  silicon photonic crystals for manipulating near-field thermal radiation},}\
  }\href {\doibase 10.1016/j.jqsrt.2017.09.002} {\bibfield  {journal} {\bibinfo
   {journal} {J. Quant. Spectrosc. Radiat. Transf.}\ }\textbf {\bibinfo
  {volume} {204}},\ \bibinfo {pages} {56--62} (\bibinfo {year}
  {2018})}\BibitemShut {NoStop}%
\bibitem [{\citenamefont {Latella}\ \emph {et~al.}(2021)\citenamefont
  {Latella}, \citenamefont {Biehs},\ and\ \citenamefont
  {Ben-Abdallah}}]{latella2021smart}%
  \BibitemOpen
  \bibfield  {author} {\bibinfo {author} {\bibfnamefont {Ivan}\ \bibnamefont
  {Latella}}, \bibinfo {author} {\bibfnamefont {Svend-Age}\ \bibnamefont
  {Biehs}}, \ and\ \bibinfo {author} {\bibfnamefont {Philippe}\ \bibnamefont
  {Ben-Abdallah}},\ }\bibfield  {title} {\enquote {\bibinfo {title} {Smart
  thermal management with near-field thermal radiation},}\ }\href {\doibase
  10.1364/OE.433539} {\bibfield  {journal} {\bibinfo  {journal} {Opt. Express}\
  }\textbf {\bibinfo {volume} {29}},\ \bibinfo {pages} {24816--24833} (\bibinfo
  {year} {2021})}\BibitemShut {NoStop}%
\bibitem [{\citenamefont {Song}\ \emph {et~al.}(2015)\citenamefont {Song},
  \citenamefont {Fiorino}, \citenamefont {Meyhofer},\ and\ \citenamefont
  {Reddy}}]{song2015near}%
  \BibitemOpen
  \bibfield  {author} {\bibinfo {author} {\bibfnamefont {Bai}\ \bibnamefont
  {Song}}, \bibinfo {author} {\bibfnamefont {Anthony}\ \bibnamefont {Fiorino}},
  \bibinfo {author} {\bibfnamefont {Edgar}\ \bibnamefont {Meyhofer}}, \ and\
  \bibinfo {author} {\bibfnamefont {Pramod}\ \bibnamefont {Reddy}},\ }\bibfield
   {title} {\enquote {\bibinfo {title} {Near-field radiative thermal transport:
  {F}rom theory to experiment},}\ }\href {\doibase 10.1063/1.4919048}
  {\bibfield  {journal} {\bibinfo  {journal} {AIP Adv.}\ }\textbf {\bibinfo
  {volume} {5}},\ \bibinfo {pages} {053503} (\bibinfo {year}
  {2015})}\BibitemShut {NoStop}%
\bibitem [{\citenamefont {Planck}(1901)}]{Planck1901}%
  \BibitemOpen
  \bibfield  {author} {\bibinfo {author} {\bibfnamefont {Max}\ \bibnamefont
  {Planck}},\ }\bibfield  {title} {\enquote {\bibinfo {title} {Ueber das
  {G}esetz der {E}nergieverteilung im {N}ormalspectrum},}\ }\href {\doibase
  https://doi.org/10.1002/andp.19013090310} {\bibfield  {journal} {\bibinfo
  {journal} {Ann. Phys.}\ }\textbf {\bibinfo {volume} {309}},\ \bibinfo {pages}
  {553--563} (\bibinfo {year} {1901})}\BibitemShut {NoStop}%
\bibitem [{\citenamefont {Bergman}(2011)}]{bergman2011fundamentals}%
  \BibitemOpen
  \bibfield  {author} {\bibinfo {author} {\bibfnamefont {Theodore~L}\
  \bibnamefont {Bergman}},\ }\href@noop {} {\emph {\bibinfo {title}
  {Fundamentals of heat and mass transfer}}}\ (\bibinfo  {publisher} {John
  Wiley \& Sons},\ \bibinfo {year} {2011})\BibitemShut {NoStop}%
\bibitem [{\citenamefont {Biehs}\ \emph {et~al.}(2021)\citenamefont {Biehs},
  \citenamefont {Messina}, \citenamefont {Venkataram}, \citenamefont
  {Rodriguez}, \citenamefont {Cuevas},\ and\ \citenamefont
  {Ben-Abdallah}}]{Biehs2021Near}%
  \BibitemOpen
  \bibfield  {author} {\bibinfo {author} {\bibfnamefont {S.-A.}\ \bibnamefont
  {Biehs}}, \bibinfo {author} {\bibfnamefont {R.}~\bibnamefont {Messina}},
  \bibinfo {author} {\bibfnamefont {P.~S.}\ \bibnamefont {Venkataram}},
  \bibinfo {author} {\bibfnamefont {A.~W.}\ \bibnamefont {Rodriguez}}, \bibinfo
  {author} {\bibfnamefont {J.~C.}\ \bibnamefont {Cuevas}}, \ and\ \bibinfo
  {author} {\bibfnamefont {P.}~\bibnamefont {Ben-Abdallah}},\ }\bibfield
  {title} {\enquote {\bibinfo {title} {Near-field radiative heat transfer in
  many-body systems},}\ }\href {\doibase 10.1103/RevModPhys.93.025009}
  {\bibfield  {journal} {\bibinfo  {journal} {Rev. Mod. Phys.}\ }\textbf
  {\bibinfo {volume} {93}},\ \bibinfo {pages} {025009} (\bibinfo {year}
  {2021})}\BibitemShut {NoStop}%
\bibitem [{\citenamefont {Rousseau}\ \emph {et~al.}(2009)\citenamefont
  {Rousseau}, \citenamefont {Siria}, \citenamefont {Jourdan}, \citenamefont
  {Volz}, \citenamefont {Comin}, \citenamefont {Chevrier},\ and\ \citenamefont
  {Greffet}}]{Rousseau2009Radiative}%
  \BibitemOpen
  \bibfield  {author} {\bibinfo {author} {\bibfnamefont {Emmanuel}\
  \bibnamefont {Rousseau}}, \bibinfo {author} {\bibfnamefont {Alessandro}\
  \bibnamefont {Siria}}, \bibinfo {author} {\bibfnamefont {Guillaume}\
  \bibnamefont {Jourdan}}, \bibinfo {author} {\bibfnamefont {Sebastian}\
  \bibnamefont {Volz}}, \bibinfo {author} {\bibfnamefont {Fabio}\ \bibnamefont
  {Comin}}, \bibinfo {author} {\bibfnamefont {Jo{\"e}l}\ \bibnamefont
  {Chevrier}}, \ and\ \bibinfo {author} {\bibfnamefont {Jean-Jacques}\
  \bibnamefont {Greffet}},\ }\bibfield  {title} {\enquote {\bibinfo {title}
  {Radiative heat transfer at the nanoscale},}\ }\href {\doibase
  10.1038/nphoton.2009.144} {\bibfield  {journal} {\bibinfo  {journal} {Nat.
  Photonics}\ }\textbf {\bibinfo {volume} {3}},\ \bibinfo {pages} {514--517}
  (\bibinfo {year} {2009})}\BibitemShut {NoStop}%
\bibitem [{\citenamefont {Lim}\ \emph {et~al.}(2018)\citenamefont {Lim},
  \citenamefont {Song}, \citenamefont {Lee},\ and\ \citenamefont
  {Lee}}]{lim2018tailoring}%
  \BibitemOpen
  \bibfield  {author} {\bibinfo {author} {\bibfnamefont {Mikyung}\ \bibnamefont
  {Lim}}, \bibinfo {author} {\bibfnamefont {Jaeman}\ \bibnamefont {Song}},
  \bibinfo {author} {\bibfnamefont {Seung~S}\ \bibnamefont {Lee}}, \ and\
  \bibinfo {author} {\bibfnamefont {Bong~Jae}\ \bibnamefont {Lee}},\ }\bibfield
   {title} {\enquote {\bibinfo {title} {Tailoring near-field thermal radiation
  between metallo-dielectric multilayers using coupled surface plasmon
  polaritons},}\ }\href {\doibase 10.1038/s41467-018-06795-w} {\bibfield
  {journal} {\bibinfo  {journal} {Nat. Commun.}\ }\textbf {\bibinfo {volume}
  {9}},\ \bibinfo {pages} {4302} (\bibinfo {year} {2018})}\BibitemShut
  {NoStop}%
\bibitem [{\citenamefont {Shen}\ \emph {et~al.}(2009)\citenamefont {Shen},
  \citenamefont {Narayanaswamy},\ and\ \citenamefont {Chen}}]{shen2009surface}%
  \BibitemOpen
  \bibfield  {author} {\bibinfo {author} {\bibfnamefont {Sheng}\ \bibnamefont
  {Shen}}, \bibinfo {author} {\bibfnamefont {Arvind}\ \bibnamefont
  {Narayanaswamy}}, \ and\ \bibinfo {author} {\bibfnamefont {Gang}\
  \bibnamefont {Chen}},\ }\bibfield  {title} {\enquote {\bibinfo {title}
  {Surface phonon polaritons mediated energy transfer between nanoscale
  gaps},}\ }\href {\doibase 10.1021/nl901208v} {\bibfield  {journal} {\bibinfo
  {journal} {Nano Lett.}\ }\textbf {\bibinfo {volume} {9}},\ \bibinfo {pages}
  {2909--2913} (\bibinfo {year} {2009})}\BibitemShut {NoStop}%
\bibitem [{\citenamefont {Zhao}\ \emph {et~al.}(2017)\citenamefont {Zhao},
  \citenamefont {Guizal}, \citenamefont {Zhang}, \citenamefont {Fan},\ and\
  \citenamefont {Antezza}}]{Zhao2017near}%
  \BibitemOpen
  \bibfield  {author} {\bibinfo {author} {\bibfnamefont {Bo}~\bibnamefont
  {Zhao}}, \bibinfo {author} {\bibfnamefont {Brahim}\ \bibnamefont {Guizal}},
  \bibinfo {author} {\bibfnamefont {Zhuomin~M.}\ \bibnamefont {Zhang}},
  \bibinfo {author} {\bibfnamefont {Shanhui}\ \bibnamefont {Fan}}, \ and\
  \bibinfo {author} {\bibfnamefont {Mauro}\ \bibnamefont {Antezza}},\
  }\bibfield  {title} {\enquote {\bibinfo {title} {Near-field heat transfer
  between graphene/h{BN} multilayers},}\ }\href {\doibase
  10.1103/PhysRevB.95.245437} {\bibfield  {journal} {\bibinfo  {journal} {Phys.
  Rev. B}\ }\textbf {\bibinfo {volume} {95}},\ \bibinfo {pages} {245437}
  (\bibinfo {year} {2017})}\BibitemShut {NoStop}%
\bibitem [{\citenamefont {Zhang}(2020)}]{Zhang_book}%
  \BibitemOpen
  \bibfield  {author} {\bibinfo {author} {\bibfnamefont {Zhuomin~M.}\
  \bibnamefont {Zhang}},\ }\href {https://doi.org/10.1007/978-3-030-45039-7}
  {\emph {\bibinfo {title} {Nano/Microscale Heat Transfer}}}\ (\bibinfo
  {publisher} {Springer, Cham},\ \bibinfo {year} {2020})\BibitemShut {NoStop}%
\bibitem [{\citenamefont {Messina}\ and\ \citenamefont
  {Ben-Abdallah}(2013)}]{messina2013graphene}%
  \BibitemOpen
  \bibfield  {author} {\bibinfo {author} {\bibfnamefont {Riccardo}\
  \bibnamefont {Messina}}\ and\ \bibinfo {author} {\bibfnamefont {Philippe}\
  \bibnamefont {Ben-Abdallah}},\ }\bibfield  {title} {\enquote {\bibinfo
  {title} {Graphene-based photovoltaic cells for near-field thermal energy
  conversion},}\ }\href {\doibase 10.1038/srep01383} {\bibfield  {journal}
  {\bibinfo  {journal} {Sci. Rep.}\ }\textbf {\bibinfo {volume} {3}},\ \bibinfo
  {pages} {1383} (\bibinfo {year} {2013})}\BibitemShut {NoStop}%
\bibitem [{\citenamefont {Park}\ \emph {et~al.}(2008)\citenamefont {Park},
  \citenamefont {Basu}, \citenamefont {King},\ and\ \citenamefont
  {Zhang}}]{PARK2008305}%
  \BibitemOpen
  \bibfield  {author} {\bibinfo {author} {\bibfnamefont {K.}~\bibnamefont
  {Park}}, \bibinfo {author} {\bibfnamefont {S.}~\bibnamefont {Basu}}, \bibinfo
  {author} {\bibfnamefont {W.P.}\ \bibnamefont {King}}, \ and\ \bibinfo
  {author} {\bibfnamefont {Z.M.}\ \bibnamefont {Zhang}},\ }\bibfield  {title}
  {\enquote {\bibinfo {title} {Performance analysis of near-field
  thermophotovoltaic devices considering absorption distribution},}\ }\href
  {\doibase https://doi.org/10.1016/j.jqsrt.2007.08.022} {\bibfield  {journal}
  {\bibinfo  {journal} {J. Quant. Spectrosc. Radiat. Transf.}\ }\textbf
  {\bibinfo {volume} {109}},\ \bibinfo {pages} {305--316} (\bibinfo {year}
  {2008})}\BibitemShut {NoStop}%
\bibitem [{\citenamefont {Zhu}\ \emph {et~al.}(2013)\citenamefont {Zhu},
  \citenamefont {Otey},\ and\ \citenamefont {Fan}}]{Linxiao2013}%
  \BibitemOpen
  \bibfield  {author} {\bibinfo {author} {\bibfnamefont {Linxiao}\ \bibnamefont
  {Zhu}}, \bibinfo {author} {\bibfnamefont {Clayton~R.}\ \bibnamefont {Otey}},
  \ and\ \bibinfo {author} {\bibfnamefont {Shanhui}\ \bibnamefont {Fan}},\
  }\bibfield  {title} {\enquote {\bibinfo {title} {Ultrahigh-contrast and
  large-bandwidth thermal rectification in near-field electromagnetic thermal
  transfer between nanoparticles},}\ }\href {\doibase
  10.1103/PhysRevB.88.184301} {\bibfield  {journal} {\bibinfo  {journal} {Phys.
  Rev. B}\ }\textbf {\bibinfo {volume} {88}},\ \bibinfo {pages} {184301}
  (\bibinfo {year} {2013})}\BibitemShut {NoStop}%
\bibitem [{\citenamefont {Chen}\ \emph {et~al.}(2016)\citenamefont {Chen},
  \citenamefont {Santhanam},\ and\ \citenamefont {Fan}}]{Kaifeng2016}%
  \BibitemOpen
  \bibfield  {author} {\bibinfo {author} {\bibfnamefont {Kaifeng}\ \bibnamefont
  {Chen}}, \bibinfo {author} {\bibfnamefont {Parthiban}\ \bibnamefont
  {Santhanam}}, \ and\ \bibinfo {author} {\bibfnamefont {Shanhui}\ \bibnamefont
  {Fan}},\ }\bibfield  {title} {\enquote {\bibinfo {title} {Near-field enhanced
  negative luminescent refrigeration},}\ }\href {\doibase
  10.1103/PhysRevApplied.6.024014} {\bibfield  {journal} {\bibinfo  {journal}
  {Phys. Rev. Appl.}\ }\textbf {\bibinfo {volume} {6}},\ \bibinfo {pages}
  {024014} (\bibinfo {year} {2016})}\BibitemShut {NoStop}%
\bibitem [{\citenamefont {Ben-Abdallah}\ and\ \citenamefont
  {Biehs}(2014)}]{Abdallah2014}%
  \BibitemOpen
  \bibfield  {author} {\bibinfo {author} {\bibfnamefont {Philippe}\
  \bibnamefont {Ben-Abdallah}}\ and\ \bibinfo {author} {\bibfnamefont
  {Svend-Age}\ \bibnamefont {Biehs}},\ }\bibfield  {title} {\enquote {\bibinfo
  {title} {Near-field thermal transistor},}\ }\href {\doibase
  10.1103/PhysRevLett.112.044301} {\bibfield  {journal} {\bibinfo  {journal}
  {Phys. Rev. Lett.}\ }\textbf {\bibinfo {volume} {112}},\ \bibinfo {pages}
  {044301} (\bibinfo {year} {2014})}\BibitemShut {NoStop}%
\bibitem [{\citenamefont {Cravalho}\ \emph {et~al.}(1967)\citenamefont
  {Cravalho}, \citenamefont {Tien},\ and\ \citenamefont
  {Caren}}]{cravalho1967effect}%
  \BibitemOpen
  \bibfield  {author} {\bibinfo {author} {\bibfnamefont {Ernest~G}\
  \bibnamefont {Cravalho}}, \bibinfo {author} {\bibfnamefont {C.~L.}\
  \bibnamefont {Tien}}, \ and\ \bibinfo {author} {\bibfnamefont {R.~P.}\
  \bibnamefont {Caren}},\ }\bibfield  {title} {\enquote {\bibinfo {title}
  {Effect of small spacings on radiative transfer between two dielectrics},}\
  }\href {\doibase 10.1115/1.3614396} {\bibfield  {journal} {\bibinfo
  {journal} {J. Heat Transfer.}\ }\textbf {\bibinfo {volume} {89}},\ \bibinfo
  {pages} {351–358} (\bibinfo {year} {1967})}\BibitemShut {NoStop}%
\bibitem [{\citenamefont {Gallinet}\ and\ \citenamefont
  {Martin}(2011)}]{Gallinet11}%
  \BibitemOpen
  \bibfield  {author} {\bibinfo {author} {\bibfnamefont {Benjamin}\
  \bibnamefont {Gallinet}}\ and\ \bibinfo {author} {\bibfnamefont {Olivier
  J.~F.}\ \bibnamefont {Martin}},\ }\bibfield  {title} {\enquote {\bibinfo
  {title} {Relation between near-field and far-field properties of plasmonic
  {F}ano resonances},}\ }\href {\doibase 10.1364/OE.19.022167} {\bibfield
  {journal} {\bibinfo  {journal} {Opt. Express}\ }\textbf {\bibinfo {volume}
  {19}},\ \bibinfo {pages} {22167--22175} (\bibinfo {year} {2011})}\BibitemShut
  {NoStop}%
\bibitem [{\citenamefont {Odebowale}\ \emph {et~al.}(2025)\citenamefont
  {Odebowale}, \citenamefont {Berhe}, \citenamefont {Ogundare}, \citenamefont
  {Abdo}, \citenamefont {Abdulghani}, \citenamefont {Hattori},\ and\
  \citenamefont {Miroshnichenko}}]{Odebowale2421051}%
  \BibitemOpen
  \bibfield  {author} {\bibinfo {author} {\bibfnamefont {Ambali~Alade}\
  \bibnamefont {Odebowale}}, \bibinfo {author} {\bibfnamefont {Andergachew}\
  \bibnamefont {Berhe}}, \bibinfo {author} {\bibfnamefont {Rasheed~T}\
  \bibnamefont {Ogundare}}, \bibinfo {author} {\bibfnamefont {Salah}\
  \bibnamefont {Abdo}}, \bibinfo {author} {\bibfnamefont {Amer}\ \bibnamefont
  {Abdulghani}}, \bibinfo {author} {\bibfnamefont {Haroldo~T}\ \bibnamefont
  {Hattori}}, \ and\ \bibinfo {author} {\bibfnamefont {Andrey~E}\ \bibnamefont
  {Miroshnichenko}},\ }\bibfield  {title} {\enquote {\bibinfo {title} {Advances
  in radiative heat transfer: Bridging far-field fundamentals and emerging
  near-field innovations},}\ }\href {\doibase 10.1002/adfm.202421051}
  {\bibfield  {journal} {\bibinfo  {journal} {Adv. Funct. Mater.}\ ,\ \bibinfo
  {pages} {2421051}} (\bibinfo {year} {2025})}\BibitemShut {NoStop}%
\bibitem [{\citenamefont {Tachikawa}\ \emph {et~al.}(2024)\citenamefont
  {Tachikawa}, \citenamefont {Ordonez-Miranda}, \citenamefont {Jalabert},
  \citenamefont {Wu}, \citenamefont {Anufriev}, \citenamefont {Guo},
  \citenamefont {Kim}, \citenamefont {Fujita}, \citenamefont {Volz},\ and\
  \citenamefont {Nomura}}]{Tachikawa2024}%
  \BibitemOpen
  \bibfield  {author} {\bibinfo {author} {\bibfnamefont {Saeko}\ \bibnamefont
  {Tachikawa}}, \bibinfo {author} {\bibfnamefont {Jose}\ \bibnamefont
  {Ordonez-Miranda}}, \bibinfo {author} {\bibfnamefont {Laurent}\ \bibnamefont
  {Jalabert}}, \bibinfo {author} {\bibfnamefont {Yunhui}\ \bibnamefont {Wu}},
  \bibinfo {author} {\bibfnamefont {Roman}\ \bibnamefont {Anufriev}}, \bibinfo
  {author} {\bibfnamefont {Yangyu}\ \bibnamefont {Guo}}, \bibinfo {author}
  {\bibfnamefont {Byunggi}\ \bibnamefont {Kim}}, \bibinfo {author}
  {\bibfnamefont {Hiroyuki}\ \bibnamefont {Fujita}}, \bibinfo {author}
  {\bibfnamefont {Sebastian}\ \bibnamefont {Volz}}, \ and\ \bibinfo {author}
  {\bibfnamefont {Masahiro}\ \bibnamefont {Nomura}},\ }\bibfield  {title}
  {\enquote {\bibinfo {title} {Enhanced {F}ar-{F}ield {T}hermal {R}adiation
  through a {P}olaritonic {W}aveguide},}\ }\href {\doibase
  10.1103/PhysRevLett.132.186904} {\bibfield  {journal} {\bibinfo  {journal}
  {Phys. Rev. Lett.}\ }\textbf {\bibinfo {volume} {132}},\ \bibinfo {pages}
  {186904} (\bibinfo {year} {2024})}\BibitemShut {NoStop}%
\bibitem [{\citenamefont {Greffet}\ \emph {et~al.}(2002)\citenamefont
  {Greffet}, \citenamefont {Carminati}, \citenamefont {Joulain}, \citenamefont
  {Mulet}, \citenamefont {Mainguy},\ and\ \citenamefont
  {Chen}}]{greffet2002coherent}%
  \BibitemOpen
  \bibfield  {author} {\bibinfo {author} {\bibfnamefont {Jean-Jacques}\
  \bibnamefont {Greffet}}, \bibinfo {author} {\bibfnamefont {R{\'e}mi}\
  \bibnamefont {Carminati}}, \bibinfo {author} {\bibfnamefont {Karl}\
  \bibnamefont {Joulain}}, \bibinfo {author} {\bibfnamefont {Jean-Philippe}\
  \bibnamefont {Mulet}}, \bibinfo {author} {\bibfnamefont {St{\'e}phane}\
  \bibnamefont {Mainguy}}, \ and\ \bibinfo {author} {\bibfnamefont {Yong}\
  \bibnamefont {Chen}},\ }\bibfield  {title} {\enquote {\bibinfo {title}
  {Coherent emission of light by thermal sources},}\ }\href {\doibase
  10.1038/416061a} {\bibfield  {journal} {\bibinfo  {journal} {Nature}\
  }\textbf {\bibinfo {volume} {416}},\ \bibinfo {pages} {61--64} (\bibinfo
  {year} {2002})}\BibitemShut {NoStop}%
\bibitem [{\citenamefont {Ding}\ \emph {et~al.}(2016)\citenamefont {Ding},
  \citenamefont {Kim},\ and\ \citenamefont {Minnich}}]{ding2016active1}%
  \BibitemOpen
  \bibfield  {author} {\bibinfo {author} {\bibfnamefont {D}~\bibnamefont
  {Ding}}, \bibinfo {author} {\bibfnamefont {T}~\bibnamefont {Kim}}, \ and\
  \bibinfo {author} {\bibfnamefont {A.~J.}\ \bibnamefont {Minnich}},\
  }\bibfield  {title} {\enquote {\bibinfo {title} {Active thermal extraction
  and temperature sensing of near-field thermal radiation},}\ }\href {\doibase
  10.1038/srep32744} {\bibfield  {journal} {\bibinfo  {journal} {Sci. Rep.}\
  }\textbf {\bibinfo {volume} {6}},\ \bibinfo {pages} {32744} (\bibinfo {year}
  {2016})}\BibitemShut {NoStop}%
\bibitem [{\citenamefont {Schuller}\ \emph {et~al.}(2009)\citenamefont
  {Schuller}, \citenamefont {Taubner},\ and\ \citenamefont
  {Brongersma}}]{schuller2009optical}%
  \BibitemOpen
  \bibfield  {author} {\bibinfo {author} {\bibfnamefont {Jon~A}\ \bibnamefont
  {Schuller}}, \bibinfo {author} {\bibfnamefont {Thomas}\ \bibnamefont
  {Taubner}}, \ and\ \bibinfo {author} {\bibfnamefont {Mark~L}\ \bibnamefont
  {Brongersma}},\ }\bibfield  {title} {\enquote {\bibinfo {title} {Optical
  antenna thermal emitters},}\ }\href {\doibase 10.1038/nphoton.2009.188}
  {\bibfield  {journal} {\bibinfo  {journal} {Nat. Photonics}\ }\textbf
  {\bibinfo {volume} {3}},\ \bibinfo {pages} {658--661} (\bibinfo {year}
  {2009})}\BibitemShut {NoStop}%
\bibitem [{\citenamefont {Yu}\ \emph {et~al.}(2013)\citenamefont {Yu},
  \citenamefont {Sergeant}, \citenamefont {Skauli}, \citenamefont {Zhang},
  \citenamefont {Wang},\ and\ \citenamefont {Fan}}]{yu2013enhancing}%
  \BibitemOpen
  \bibfield  {author} {\bibinfo {author} {\bibfnamefont {Zongfu}\ \bibnamefont
  {Yu}}, \bibinfo {author} {\bibfnamefont {Nicholas~P}\ \bibnamefont
  {Sergeant}}, \bibinfo {author} {\bibfnamefont {Torbj{\o}rn}\ \bibnamefont
  {Skauli}}, \bibinfo {author} {\bibfnamefont {Gang}\ \bibnamefont {Zhang}},
  \bibinfo {author} {\bibfnamefont {Hailiang}\ \bibnamefont {Wang}}, \ and\
  \bibinfo {author} {\bibfnamefont {Shanhui}\ \bibnamefont {Fan}},\ }\bibfield
  {title} {\enquote {\bibinfo {title} {Enhancing far-field thermal emission
  with thermal extraction},}\ }\href {\doibase 10.1038/ncomms2765} {\bibfield
  {journal} {\bibinfo  {journal} {Nat. Commun.}\ }\textbf {\bibinfo {volume}
  {4}},\ \bibinfo {pages} {1730} (\bibinfo {year} {2013})}\BibitemShut
  {NoStop}%
\bibitem [{\citenamefont {Taravati}\ and\ \citenamefont
  {Eleftheriades}(2021)}]{taravati2021pure}%
  \BibitemOpen
  \bibfield  {author} {\bibinfo {author} {\bibfnamefont {Sajjad}\ \bibnamefont
  {Taravati}}\ and\ \bibinfo {author} {\bibfnamefont {George~V}\ \bibnamefont
  {Eleftheriades}},\ }\bibfield  {title} {\enquote {\bibinfo {title} {Pure and
  {L}inear {F}requency-{C}onversion {T}emporal {M}etasurface},}\ }\href
  {\doibase 10.1103/PhysRevApplied.15.064011} {\bibfield  {journal} {\bibinfo
  {journal} {Phys. Rev. Appl.}\ }\textbf {\bibinfo {volume} {15}},\ \bibinfo
  {pages} {064011} (\bibinfo {year} {2021})}\BibitemShut {NoStop}%
\bibitem [{\citenamefont {Shcherbakov}\ \emph {et~al.}(2019)\citenamefont
  {Shcherbakov}, \citenamefont {Werner}, \citenamefont {Fan}, \citenamefont
  {Talisa}, \citenamefont {Chowdhury},\ and\ \citenamefont
  {Shvets}}]{shcherbakov2019photon}%
  \BibitemOpen
  \bibfield  {author} {\bibinfo {author} {\bibfnamefont {Maxim~R}\ \bibnamefont
  {Shcherbakov}}, \bibinfo {author} {\bibfnamefont {Kevin}\ \bibnamefont
  {Werner}}, \bibinfo {author} {\bibfnamefont {Zhiyuan}\ \bibnamefont {Fan}},
  \bibinfo {author} {\bibfnamefont {Noah}\ \bibnamefont {Talisa}}, \bibinfo
  {author} {\bibfnamefont {Enam}\ \bibnamefont {Chowdhury}}, \ and\ \bibinfo
  {author} {\bibfnamefont {Gennady}\ \bibnamefont {Shvets}},\ }\bibfield
  {title} {\enquote {\bibinfo {title} {Photon acceleration and tunable
  broadband harmonics generation in nonlinear time-dependent metasurfaces},}\
  }\href {\doibase 10.1038/s41467-019-09313-8} {\bibfield  {journal} {\bibinfo
  {journal} {Nat. Commun.}\ }\textbf {\bibinfo {volume} {10}},\ \bibinfo
  {pages} {1345} (\bibinfo {year} {2019})}\BibitemShut {NoStop}%
\bibitem [{\citenamefont {Ramaccia}\ \emph {et~al.}(2019)\citenamefont
  {Ramaccia}, \citenamefont {Sounas}, \citenamefont {Alù}, \citenamefont
  {Toscano},\ and\ \citenamefont {Bilotti}}]{ramaccia2019phase}%
  \BibitemOpen
  \bibfield  {author} {\bibinfo {author} {\bibfnamefont {Davide}\ \bibnamefont
  {Ramaccia}}, \bibinfo {author} {\bibfnamefont {Dimitrios~L.}\ \bibnamefont
  {Sounas}}, \bibinfo {author} {\bibfnamefont {Andrea}\ \bibnamefont {Alù}},
  \bibinfo {author} {\bibfnamefont {Alessandro}\ \bibnamefont {Toscano}}, \
  and\ \bibinfo {author} {\bibfnamefont {Filiberto}\ \bibnamefont {Bilotti}},\
  }\bibfield  {title} {\enquote {\bibinfo {title} {{P}hase-{I}nduced
  {F}requency {C}onversion and {D}oppler {E}ffect {W}ith {T}ime-{M}odulated
  {M}etasurfaces},}\ }\href {\doibase 10.1109/TAP.2019.2952469} {\bibfield
  {journal} {\bibinfo  {journal} {IEEE Trans. Antennas Propag.}\ }\textbf
  {\bibinfo {volume} {68}},\ \bibinfo {pages} {1607--1617} (\bibinfo {year}
  {2019})}\BibitemShut {NoStop}%
\bibitem [{\citenamefont {Zhou}\ \emph {et~al.}(2020)\citenamefont {Zhou},
  \citenamefont {Alam}, \citenamefont {Karimi}, \citenamefont {Upham},
  \citenamefont {Reshef}, \citenamefont {Liu}, \citenamefont {Willner},\ and\
  \citenamefont {Boyd}}]{zhou2020broadband}%
  \BibitemOpen
  \bibfield  {author} {\bibinfo {author} {\bibfnamefont {Yiyu}\ \bibnamefont
  {Zhou}}, \bibinfo {author} {\bibfnamefont {M~Zahirul}\ \bibnamefont {Alam}},
  \bibinfo {author} {\bibfnamefont {Mohammad}\ \bibnamefont {Karimi}}, \bibinfo
  {author} {\bibfnamefont {Jeremy}\ \bibnamefont {Upham}}, \bibinfo {author}
  {\bibfnamefont {Orad}\ \bibnamefont {Reshef}}, \bibinfo {author}
  {\bibfnamefont {Cong}\ \bibnamefont {Liu}}, \bibinfo {author} {\bibfnamefont
  {Alan~E}\ \bibnamefont {Willner}}, \ and\ \bibinfo {author} {\bibfnamefont
  {Robert~W}\ \bibnamefont {Boyd}},\ }\bibfield  {title} {\enquote {\bibinfo
  {title} {Broadband frequency translation through time refraction in an
  epsilon-near-zero material},}\ }\href {\doibase 10.1038/s41467-020-15682-2}
  {\bibfield  {journal} {\bibinfo  {journal} {Nat. Commun.}\ }\textbf {\bibinfo
  {volume} {11}},\ \bibinfo {pages} {2180} (\bibinfo {year}
  {2020})}\BibitemShut {NoStop}%
\bibitem [{\citenamefont {Pang}\ \emph {et~al.}(2021)\citenamefont {Pang},
  \citenamefont {Alam}, \citenamefont {Zhou}, \citenamefont {Liu},
  \citenamefont {Reshef}, \citenamefont {Manukyan}, \citenamefont {Voegtle},
  \citenamefont {Pennathur}, \citenamefont {Tseng}, \citenamefont {Su} \emph
  {et~al.}}]{pang2021adiabatic}%
  \BibitemOpen
  \bibfield  {author} {\bibinfo {author} {\bibfnamefont {Kai}\ \bibnamefont
  {Pang}}, \bibinfo {author} {\bibfnamefont {M~Zahirul}\ \bibnamefont {Alam}},
  \bibinfo {author} {\bibfnamefont {Yiyu}\ \bibnamefont {Zhou}}, \bibinfo
  {author} {\bibfnamefont {Cong}\ \bibnamefont {Liu}}, \bibinfo {author}
  {\bibfnamefont {Orad}\ \bibnamefont {Reshef}}, \bibinfo {author}
  {\bibfnamefont {Karapet}\ \bibnamefont {Manukyan}}, \bibinfo {author}
  {\bibfnamefont {Matt}\ \bibnamefont {Voegtle}}, \bibinfo {author}
  {\bibfnamefont {Anuj}\ \bibnamefont {Pennathur}}, \bibinfo {author}
  {\bibfnamefont {Cindy}\ \bibnamefont {Tseng}}, \bibinfo {author}
  {\bibfnamefont {Xinzhou}\ \bibnamefont {Su}},  \emph {et~al.},\ }\bibfield
  {title} {\enquote {\bibinfo {title} {Adiabatic frequency conversion using a
  time-varying epsilon-near-zero metasurface},}\ }\href {\doibase
  10.1021/acs.nanolett.1c00550} {\bibfield  {journal} {\bibinfo  {journal}
  {Nano Lett.}\ }\textbf {\bibinfo {volume} {21}},\ \bibinfo {pages}
  {5907--5913} (\bibinfo {year} {2021})}\BibitemShut {NoStop}%
\bibitem [{\citenamefont {Sounas}\ and\ \citenamefont
  {Al{\`u}}(2017)}]{sounas2017non}%
  \BibitemOpen
  \bibfield  {author} {\bibinfo {author} {\bibfnamefont {Dimitrios~L}\
  \bibnamefont {Sounas}}\ and\ \bibinfo {author} {\bibfnamefont {Andrea}\
  \bibnamefont {Al{\`u}}},\ }\bibfield  {title} {\enquote {\bibinfo {title}
  {Non-reciprocal photonics based on time modulation},}\ }\href {\doibase
  10.1038/s41566-017-0051-x} {\bibfield  {journal} {\bibinfo  {journal} {Nat.
  Photonics}\ }\textbf {\bibinfo {volume} {11}},\ \bibinfo {pages} {774--783}
  (\bibinfo {year} {2017})}\BibitemShut {NoStop}%
\bibitem [{\citenamefont {Cardin}\ \emph {et~al.}(2020)\citenamefont {Cardin},
  \citenamefont {Silva}, \citenamefont {Vardeny}, \citenamefont {Padilla},
  \citenamefont {Saxena}, \citenamefont {Taylor}, \citenamefont {Kort-Kamp},
  \citenamefont {Chen}, \citenamefont {Dalvit},\ and\ \citenamefont
  {Azad}}]{cardin2020surface}%
  \BibitemOpen
  \bibfield  {author} {\bibinfo {author} {\bibfnamefont {Andrew~E}\
  \bibnamefont {Cardin}}, \bibinfo {author} {\bibfnamefont {Sinhara~R}\
  \bibnamefont {Silva}}, \bibinfo {author} {\bibfnamefont {Shai~R}\
  \bibnamefont {Vardeny}}, \bibinfo {author} {\bibfnamefont {Willie~J}\
  \bibnamefont {Padilla}}, \bibinfo {author} {\bibfnamefont {Avadh}\
  \bibnamefont {Saxena}}, \bibinfo {author} {\bibfnamefont {Antoinette~J}\
  \bibnamefont {Taylor}}, \bibinfo {author} {\bibfnamefont {Wilton~JM}\
  \bibnamefont {Kort-Kamp}}, \bibinfo {author} {\bibfnamefont {Hou-Tong}\
  \bibnamefont {Chen}}, \bibinfo {author} {\bibfnamefont {Diego~AR}\
  \bibnamefont {Dalvit}}, \ and\ \bibinfo {author} {\bibfnamefont {Abul~K}\
  \bibnamefont {Azad}},\ }\bibfield  {title} {\enquote {\bibinfo {title}
  {Surface-wave-assisted nonreciprocity in spatio-temporally modulated
  metasurfaces},}\ }\href {\doibase 10.1038/s41467-020-15273-1} {\bibfield
  {journal} {\bibinfo  {journal} {Nat. Commun.}\ }\textbf {\bibinfo {volume}
  {11}},\ \bibinfo {pages} {1469} (\bibinfo {year} {2020})}\BibitemShut
  {NoStop}%
\bibitem [{\citenamefont {Barati~Sedeh}\ \emph {et~al.}(2022)\citenamefont
  {Barati~Sedeh}, \citenamefont {Mohammadi~Dinani},\ and\ \citenamefont
  {Mosallaei}}]{barati2022optical}%
  \BibitemOpen
  \bibfield  {author} {\bibinfo {author} {\bibfnamefont {Hooman}\ \bibnamefont
  {Barati~Sedeh}}, \bibinfo {author} {\bibfnamefont {Hediyeh}\ \bibnamefont
  {Mohammadi~Dinani}}, \ and\ \bibinfo {author} {\bibfnamefont {Hossein}\
  \bibnamefont {Mosallaei}},\ }\bibfield  {title} {\enquote {\bibinfo {title}
  {Optical nonreciprocity via transmissive time-modulated metasurfaces},}\
  }\href {\doibase doi:10.1515/nanoph-2022-0373} {\bibfield  {journal}
  {\bibinfo  {journal} {Nanophotonics}\ }\textbf {\bibinfo {volume} {11}},\
  \bibinfo {pages} {4135--4148} (\bibinfo {year} {2022})}\BibitemShut {NoStop}%
\bibitem [{\citenamefont {Galiffi}\ \emph {et~al.}(2020)\citenamefont
  {Galiffi}, \citenamefont {Wang}, \citenamefont {Lim}, \citenamefont {Pendry},
  \citenamefont {Al{\`u}},\ and\ \citenamefont {Huidobro}}]{galiffi2020wood}%
  \BibitemOpen
  \bibfield  {author} {\bibinfo {author} {\bibfnamefont {Emanuele}\
  \bibnamefont {Galiffi}}, \bibinfo {author} {\bibfnamefont {Yao-Ting}\
  \bibnamefont {Wang}}, \bibinfo {author} {\bibfnamefont {Zhen}\ \bibnamefont
  {Lim}}, \bibinfo {author} {\bibfnamefont {John~B}\ \bibnamefont {Pendry}},
  \bibinfo {author} {\bibfnamefont {Andrea}\ \bibnamefont {Al{\`u}}}, \ and\
  \bibinfo {author} {\bibfnamefont {Paloma~A}\ \bibnamefont {Huidobro}},\
  }\bibfield  {title} {\enquote {\bibinfo {title} {Wood {A}nomalies and
  {S}urface-{W}ave {E}xcitation with a {T}ime {G}rating},}\ }\href {\doibase
  10.1103/PhysRevLett.125.127403} {\bibfield  {journal} {\bibinfo  {journal}
  {Phys. Rev. Lett.}\ }\textbf {\bibinfo {volume} {125}},\ \bibinfo {pages}
  {127403} (\bibinfo {year} {2020})}\BibitemShut {NoStop}%
\bibitem [{\citenamefont {Tsai}\ \emph {et~al.}(2022)\citenamefont {Tsai},
  \citenamefont {Wang}, \citenamefont {Galiffi}, \citenamefont {Al{\`u}},\ and\
  \citenamefont {Yen}}]{tsai2022surface}%
  \BibitemOpen
  \bibfield  {author} {\bibinfo {author} {\bibfnamefont {Ya-Wen}\ \bibnamefont
  {Tsai}}, \bibinfo {author} {\bibfnamefont {Yao-Ting}\ \bibnamefont {Wang}},
  \bibinfo {author} {\bibfnamefont {Emanuele}\ \bibnamefont {Galiffi}},
  \bibinfo {author} {\bibfnamefont {Andrea}\ \bibnamefont {Al{\`u}}}, \ and\
  \bibinfo {author} {\bibfnamefont {Ta-Jen}\ \bibnamefont {Yen}},\ }\bibfield
  {title} {\enquote {\bibinfo {title} {Surface-wave coupling in double
  {F}loquet sheets supporting phased temporal {W}ood anomalies},}\ }\href
  {\doibase doi:10.1515/nanoph-2022-0253} {\bibfield  {journal} {\bibinfo
  {journal} {Nanophotonics}\ }\textbf {\bibinfo {volume} {11}},\ \bibinfo
  {pages} {3509--3517} (\bibinfo {year} {2022})}\BibitemShut {NoStop}%
\bibitem [{\citenamefont {Tang}\ and\ \citenamefont
  {Wang}(2024)}]{tang2024modulating}%
  \BibitemOpen
  \bibfield  {author} {\bibinfo {author} {\bibfnamefont {Gaomin}\ \bibnamefont
  {Tang}}\ and\ \bibinfo {author} {\bibfnamefont {Jian-Sheng}\ \bibnamefont
  {Wang}},\ }\bibfield  {title} {\enquote {\bibinfo {title} {Modulating
  near-field thermal transfer through temporal drivings: {A} quantum many-body
  theory},}\ }\href {\doibase 10.1103/PhysRevB.109.085428} {\bibfield
  {journal} {\bibinfo  {journal} {Phys. Rev. B}\ }\textbf {\bibinfo {volume}
  {109}},\ \bibinfo {pages} {085428} (\bibinfo {year} {2024})}\BibitemShut
  {NoStop}%
\bibitem [{\citenamefont {Coppens}\ and\ \citenamefont
  {Valentine}(2017)}]{coppens2017spatial}%
  \BibitemOpen
  \bibfield  {author} {\bibinfo {author} {\bibfnamefont {Zachary~J}\
  \bibnamefont {Coppens}}\ and\ \bibinfo {author} {\bibfnamefont {Jason~G}\
  \bibnamefont {Valentine}},\ }\bibfield  {title} {\enquote {\bibinfo {title}
  {Spatial and temporal modulation of thermal emission},}\ }\href {\doibase
  10.1002/adma.201701275} {\bibfield  {journal} {\bibinfo  {journal} {Adv.
  Mater.}\ }\textbf {\bibinfo {volume} {29}},\ \bibinfo {pages} {1701275}
  (\bibinfo {year} {2017})}\BibitemShut {NoStop}%
\bibitem [{\citenamefont {Buddhiraju}\ \emph {et~al.}(2020)\citenamefont
  {Buddhiraju}, \citenamefont {Li},\ and\ \citenamefont
  {Fan}}]{buddhiraju2020photonic}%
  \BibitemOpen
  \bibfield  {author} {\bibinfo {author} {\bibfnamefont {Siddharth}\
  \bibnamefont {Buddhiraju}}, \bibinfo {author} {\bibfnamefont {Wei}\
  \bibnamefont {Li}}, \ and\ \bibinfo {author} {\bibfnamefont {Shanhui}\
  \bibnamefont {Fan}},\ }\bibfield  {title} {\enquote {\bibinfo {title}
  {Photonic refrigeration from time-modulated thermal emission},}\ }\href
  {\doibase 10.1103/PhysRevLett.124.077402} {\bibfield  {journal} {\bibinfo
  {journal} {Phys. Rev. Lett.}\ }\textbf {\bibinfo {volume} {124}},\ \bibinfo
  {pages} {077402} (\bibinfo {year} {2020})}\BibitemShut {NoStop}%
\bibitem [{\citenamefont {Yu}\ and\ \citenamefont {Fan}(2024)}]{yu2024time}%
  \BibitemOpen
  \bibfield  {author} {\bibinfo {author} {\bibfnamefont {Renwen}\ \bibnamefont
  {Yu}}\ and\ \bibinfo {author} {\bibfnamefont {Shanhui}\ \bibnamefont {Fan}},\
  }\bibfield  {title} {\enquote {\bibinfo {title} {Time-modulated near-field
  radiative heat transfer},}\ }\href {\doibase 10.1073/pnas.2401514121}
  {\bibfield  {journal} {\bibinfo  {journal} {Proc. Natl. Acad. Sci.}\ }\textbf
  {\bibinfo {volume} {121}},\ \bibinfo {pages} {e2401514121} (\bibinfo {year}
  {2024})}\BibitemShut {NoStop}%
\bibitem [{\citenamefont {Ordonez-Miranda}\ \emph {et~al.}(2024)\citenamefont
  {Ordonez-Miranda}, \citenamefont {Wu}, \citenamefont {Nomura},\ and\
  \citenamefont {Volz}}]{ordonez2024far}%
  \BibitemOpen
  \bibfield  {author} {\bibinfo {author} {\bibfnamefont {Jose}\ \bibnamefont
  {Ordonez-Miranda}}, \bibinfo {author} {\bibfnamefont {Yunhui}\ \bibnamefont
  {Wu}}, \bibinfo {author} {\bibfnamefont {Masahiro}\ \bibnamefont {Nomura}}, \
  and\ \bibinfo {author} {\bibfnamefont {Sebastian}\ \bibnamefont {Volz}},\
  }\bibfield  {title} {\enquote {\bibinfo {title} {Far-field thermal radiation
  driven by the temperature oscillations of macroscopic bodies},}\ }\href
  {\doibase 10.1103/PhysRevApplied.22.054053} {\bibfield  {journal} {\bibinfo
  {journal} {Phys. Rev. Appl.}\ }\textbf {\bibinfo {volume} {22}},\ \bibinfo
  {pages} {054053} (\bibinfo {year} {2024})}\BibitemShut {NoStop}%
\bibitem [{\citenamefont {V{\'a}zquez-Lozano}\ and\ \citenamefont
  {Liberal}(2023)}]{vazquez2023incandescent}%
  \BibitemOpen
  \bibfield  {author} {\bibinfo {author} {\bibfnamefont {J~Enrique}\
  \bibnamefont {V{\'a}zquez-Lozano}}\ and\ \bibinfo {author} {\bibfnamefont
  {I{\~n}igo}\ \bibnamefont {Liberal}},\ }\bibfield  {title} {\enquote
  {\bibinfo {title} {Incandescent temporal metamaterials},}\ }\href {\doibase
  10.1038/s41467-023-40281-2} {\bibfield  {journal} {\bibinfo  {journal} {Nat.
  Commun.}\ }\textbf {\bibinfo {volume} {14}},\ \bibinfo {pages} {4606}
  (\bibinfo {year} {2023})}\BibitemShut {NoStop}%
\bibitem [{\citenamefont {Vertiz-Conde}\ \emph {et~al.}(2025)\citenamefont
  {Vertiz-Conde}, \citenamefont {Liberal},\ and\ \citenamefont {Enrique
  V{\'a}zquez-Lozano}}]{vertiz2025dispersion}%
  \BibitemOpen
  \bibfield  {author} {\bibinfo {author} {\bibfnamefont {Amaia}\ \bibnamefont
  {Vertiz-Conde}}, \bibinfo {author} {\bibfnamefont {I{\~n}igo}\ \bibnamefont
  {Liberal}}, \ and\ \bibinfo {author} {\bibfnamefont {J}~\bibnamefont {Enrique
  V{\'a}zquez-Lozano}},\ }\bibfield  {title} {\enquote {\bibinfo {title}
  {Dispersion effects in thermal emission from temporal metamaterials:
  high-frequency cutoffs},}\ }\href {\doibase 10.1364/OL.545236} {\bibfield
  {journal} {\bibinfo  {journal} {Opt. Lett.}\ }\textbf {\bibinfo {volume}
  {50}},\ \bibinfo {pages} {1097--1100} (\bibinfo {year} {2025})}\BibitemShut
  {NoStop}%
\bibitem [{\citenamefont {Liberal}\ \emph {et~al.}(2025)\citenamefont
  {Liberal}, \citenamefont {V\'{a}zquez-Lozano},\ and\ \citenamefont
  {Ganfornina-Andrades}}]{liberal2024can}%
  \BibitemOpen
  \bibfield  {author} {\bibinfo {author} {\bibfnamefont {I{\~n}igo}\
  \bibnamefont {Liberal}}, \bibinfo {author} {\bibfnamefont {J.~E.}\
  \bibnamefont {V\'{a}zquez-Lozano}}, \ and\ \bibinfo {author} {\bibfnamefont
  {Antonio}\ \bibnamefont {Ganfornina-Andrades}},\ }\bibfield  {title}
  {\enquote {\bibinfo {title} {Can thermal emission from time-varying media be
  described semiclassically?}}\ }\href {\doibase 10.1364/OME.561748} {\bibfield
   {journal} {\bibinfo  {journal} {Opt. Mater. Express}\ }\textbf {\bibinfo
  {volume} {15}},\ \bibinfo {pages} {1483--1495} (\bibinfo {year}
  {2025})}\BibitemShut {NoStop}%
\bibitem [{\citenamefont {Yu}\ and\ \citenamefont
  {Fan}(2023)}]{yu2023manipulating}%
  \BibitemOpen
  \bibfield  {author} {\bibinfo {author} {\bibfnamefont {Renwen}\ \bibnamefont
  {Yu}}\ and\ \bibinfo {author} {\bibfnamefont {Shanhui}\ \bibnamefont {Fan}},\
  }\bibfield  {title} {\enquote {\bibinfo {title} {Manipulating coherence of
  near-field thermal radiation in time-modulated systems},}\ }\href {\doibase
  10.1103/PhysRevLett.130.096902} {\bibfield  {journal} {\bibinfo  {journal}
  {Phys. Rev. Lett.}\ }\textbf {\bibinfo {volume} {130}},\ \bibinfo {pages}
  {096902} (\bibinfo {year} {2023})}\BibitemShut {NoStop}%
\bibitem [{\citenamefont {Wang}\ \emph {et~al.}(2023)\citenamefont {Wang},
  \citenamefont {Peng}, \citenamefont {Zhang}, \citenamefont {Zhang},\ and\
  \citenamefont {Zhu}}]{wang2023transport}%
  \BibitemOpen
  \bibfield  {author} {\bibinfo {author} {\bibfnamefont {Jian-Sheng}\
  \bibnamefont {Wang}}, \bibinfo {author} {\bibfnamefont {Jiebin}\ \bibnamefont
  {Peng}}, \bibinfo {author} {\bibfnamefont {Zu-Quan}\ \bibnamefont {Zhang}},
  \bibinfo {author} {\bibfnamefont {Yong-Mei}\ \bibnamefont {Zhang}}, \ and\
  \bibinfo {author} {\bibfnamefont {Tao}\ \bibnamefont {Zhu}},\ }\bibfield
  {title} {\enquote {\bibinfo {title} {Transport in electron-photon systems},}\
  }\href {\doibase /10.1007/s11467-023-1260-z} {\bibfield  {journal} {\bibinfo
  {journal} {Front. Phys.}\ }\textbf {\bibinfo {volume} {18}},\ \bibinfo
  {pages} {43602} (\bibinfo {year} {2023})}\BibitemShut {NoStop}%
\bibitem [{\citenamefont {Griffiths}(2017)}]{griffiths2023}%
  \BibitemOpen
  \bibfield  {author} {\bibinfo {author} {\bibfnamefont {David~J}\ \bibnamefont
  {Griffiths}},\ }\href@noop {} {\emph {\bibinfo {title} {Introduction to
  electrodynamics}}}\ (\bibinfo  {publisher} {Cambridge University Press},\
  \bibinfo {year} {2017})\BibitemShut {NoStop}%
\bibitem [{\citenamefont {Wang}\ \emph {et~al.}(2008)\citenamefont {Wang},
  \citenamefont {Wang},\ and\ \citenamefont {L{\"u}}}]{wang2008quantum}%
  \BibitemOpen
  \bibfield  {author} {\bibinfo {author} {\bibfnamefont {J-S}\ \bibnamefont
  {Wang}}, \bibinfo {author} {\bibfnamefont {Jian}\ \bibnamefont {Wang}}, \
  and\ \bibinfo {author} {\bibfnamefont {J.~T.}\ \bibnamefont {L{\"u}}},\
  }\bibfield  {title} {\enquote {\bibinfo {title} {Quantum thermal transport in
  nanostructures},}\ }\href {\doibase 10.1140/epjb/e2008-00195-8} {\bibfield
  {journal} {\bibinfo  {journal} {Eur. Phys. J. B}\ }\textbf {\bibinfo {volume}
  {62}},\ \bibinfo {pages} {381--404} (\bibinfo {year} {2008})}\BibitemShut
  {NoStop}%
\bibitem [{\citenamefont {Zhang}\ \emph {et~al.}(2022)\citenamefont {Zhang},
  \citenamefont {Zhu}, \citenamefont {Zhang},\ and\ \citenamefont
  {Wang}}]{zhang2022microscopic}%
  \BibitemOpen
  \bibfield  {author} {\bibinfo {author} {\bibfnamefont {Yong-Mei}\
  \bibnamefont {Zhang}}, \bibinfo {author} {\bibfnamefont {Tao}\ \bibnamefont
  {Zhu}}, \bibinfo {author} {\bibfnamefont {Zu-Quan}\ \bibnamefont {Zhang}}, \
  and\ \bibinfo {author} {\bibfnamefont {Jian-Sheng}\ \bibnamefont {Wang}},\
  }\bibfield  {title} {\enquote {\bibinfo {title} {Microscopic theory of
  photon-induced energy, momentum, and angular momentum transport in the
  nonequilibrium regime},}\ }\href {\doibase 10.1103/PhysRevB.105.205421}
  {\bibfield  {journal} {\bibinfo  {journal} {Phys. Rev. B}\ }\textbf {\bibinfo
  {volume} {105}},\ \bibinfo {pages} {205421} (\bibinfo {year}
  {2022})}\BibitemShut {NoStop}%
\bibitem [{\citenamefont {Wang}\ \emph {et~al.}(2014)\citenamefont {Wang},
  \citenamefont {Agarwalla}, \citenamefont {Li},\ and\ \citenamefont
  {Thingna}}]{wang2014nonequilibrium}%
  \BibitemOpen
  \bibfield  {author} {\bibinfo {author} {\bibfnamefont {Jian-Sheng}\
  \bibnamefont {Wang}}, \bibinfo {author} {\bibfnamefont {Bijay~Kumar}\
  \bibnamefont {Agarwalla}}, \bibinfo {author} {\bibfnamefont {Huanan}\
  \bibnamefont {Li}}, \ and\ \bibinfo {author} {\bibfnamefont {Juzar}\
  \bibnamefont {Thingna}},\ }\bibfield  {title} {\enquote {\bibinfo {title}
  {Nonequilibrium {G}reen’s function method for quantum thermal transport},}\
  }\href {\doibase 10.1007/s11467-013-0340-x} {\bibfield  {journal} {\bibinfo
  {journal} {Front. Phys.}\ }\textbf {\bibinfo {volume} {9}},\ \bibinfo {pages}
  {673--697} (\bibinfo {year} {2014})}\BibitemShut {NoStop}%
\bibitem [{sup()}]{supple}%
  \BibitemOpen
  \href@noop {} {}\bibinfo {howpublished} {See Supplemental Material for
  derivations of the free photon and retarded Green's functions, the energy
  currents, and the dispersion relations of the surface phonon polaritons,
  which includes
  Refs.~\cite{tsuji2008correlated,keller2011quantum,keldysh1965,dyson1949s,meir1992landauer,kohler2005driven}}\BibitemShut
  {NoStop}%
\bibitem [{\citenamefont {Kr\"uger}\ \emph {et~al.}(2012)\citenamefont
  {Kr\"uger}, \citenamefont {Bimonte}, \citenamefont {Emig},\ and\
  \citenamefont {Kardar}}]{Matthias2012}%
  \BibitemOpen
  \bibfield  {author} {\bibinfo {author} {\bibfnamefont {Matthias}\
  \bibnamefont {Kr\"uger}}, \bibinfo {author} {\bibfnamefont {Giuseppe}\
  \bibnamefont {Bimonte}}, \bibinfo {author} {\bibfnamefont {Thorsten}\
  \bibnamefont {Emig}}, \ and\ \bibinfo {author} {\bibfnamefont {Mehran}\
  \bibnamefont {Kardar}},\ }\bibfield  {title} {\enquote {\bibinfo {title}
  {Trace formulas for nonequilibrium {C}asimir interactions, heat radiation,
  and heat transfer for arbitrary objects},}\ }\href {\doibase
  10.1103/PhysRevB.86.115423} {\bibfield  {journal} {\bibinfo  {journal} {Phys.
  Rev. B}\ }\textbf {\bibinfo {volume} {86}},\ \bibinfo {pages} {115423}
  (\bibinfo {year} {2012})}\BibitemShut {NoStop}%
\bibitem [{\citenamefont {Joulain}\ \emph {et~al.}(2005)\citenamefont
  {Joulain}, \citenamefont {Mulet}, \citenamefont {Marquier}, \citenamefont
  {Carminati},\ and\ \citenamefont {Greffet}}]{joulain2005surface}%
  \BibitemOpen
  \bibfield  {author} {\bibinfo {author} {\bibfnamefont {Karl}\ \bibnamefont
  {Joulain}}, \bibinfo {author} {\bibfnamefont {Jean-Philippe}\ \bibnamefont
  {Mulet}}, \bibinfo {author} {\bibfnamefont {Fran{\c{c}}ois}\ \bibnamefont
  {Marquier}}, \bibinfo {author} {\bibfnamefont {R{\'e}mi}\ \bibnamefont
  {Carminati}}, \ and\ \bibinfo {author} {\bibfnamefont {Jean-Jacques}\
  \bibnamefont {Greffet}},\ }\bibfield  {title} {\enquote {\bibinfo {title}
  {Surface electromagnetic waves thermally excited: Radiative heat transfer,
  coherence properties and {C}asimir forces revisited in the near field},}\
  }\href {\doibase doi: 10.1016/j.surfrep.2004.12.002} {\bibfield  {journal}
  {\bibinfo  {journal} {Surf. Sci. Rep.}\ }\textbf {\bibinfo {volume} {57}},\
  \bibinfo {pages} {59--112} (\bibinfo {year} {2005})}\BibitemShut {NoStop}%
\bibitem [{\citenamefont {Mikhailov}\ and\ \citenamefont
  {Ziegler}(2007)}]{mikhailov2007new}%
  \BibitemOpen
  \bibfield  {author} {\bibinfo {author} {\bibfnamefont {Sergey~A}\
  \bibnamefont {Mikhailov}}\ and\ \bibinfo {author} {\bibfnamefont {Klaus}\
  \bibnamefont {Ziegler}},\ }\bibfield  {title} {\enquote {\bibinfo {title}
  {New electromagnetic mode in graphene},}\ }\href {\doibase
  10.1103/PhysRevLett.99.016803} {\bibfield  {journal} {\bibinfo  {journal}
  {Phys. Rev. Lett.}\ }\textbf {\bibinfo {volume} {99}},\ \bibinfo {pages}
  {016803} (\bibinfo {year} {2007})}\BibitemShut {NoStop}%
\bibitem [{\citenamefont {Polder}\ and\ \citenamefont
  {Van~Hove}(1971)}]{polder1971theory}%
  \BibitemOpen
  \bibfield  {author} {\bibinfo {author} {\bibfnamefont {DVHM}\ \bibnamefont
  {Polder}}\ and\ \bibinfo {author} {\bibfnamefont {M}~\bibnamefont
  {Van~Hove}},\ }\bibfield  {title} {\enquote {\bibinfo {title} {Theory of
  radiative heat transfer between closely spaced bodies},}\ }\href {\doibase
  10.1103/PhysRevB.4.3303} {\bibfield  {journal} {\bibinfo  {journal} {Phys.
  Rev. B}\ }\textbf {\bibinfo {volume} {4}},\ \bibinfo {pages} {3303} (\bibinfo
  {year} {1971})}\BibitemShut {NoStop}%
\bibitem [{\citenamefont {Pan}\ \emph {et~al.}(2025)\citenamefont {Pan},
  \citenamefont {Ren}, \citenamefont {Tang},\ and\ \citenamefont
  {Wang}}]{Pan2025Asymmetry}%
  \BibitemOpen
  \bibfield  {author} {\bibinfo {author} {\bibfnamefont {Hui}\ \bibnamefont
  {Pan}}, \bibinfo {author} {\bibfnamefont {Yuhua}\ \bibnamefont {Ren}},
  \bibinfo {author} {\bibfnamefont {Gaomin}\ \bibnamefont {Tang}}, \ and\
  \bibinfo {author} {\bibfnamefont {Jian-Sheng}\ \bibnamefont {Wang}},\
  }\bibfield  {title} {\enquote {\bibinfo {title} {Asymmetry-induced radiative
  heat transfer in {F}loquet systems},}\ }\href {\doibase 10.1103/74rq-f642}
  {\bibfield  {journal} {\bibinfo  {journal} {Phys. Rev. B}\ }\textbf {\bibinfo
  {volume} {112}},\ \bibinfo {pages} {L041401} (\bibinfo {year}
  {2025})}\BibitemShut {NoStop}%
\bibitem [{\citenamefont {Tsuji}\ \emph {et~al.}(2008)\citenamefont {Tsuji},
  \citenamefont {Oka},\ and\ \citenamefont {Aoki}}]{tsuji2008correlated}%
  \BibitemOpen
  \bibfield  {author} {\bibinfo {author} {\bibfnamefont {Naoto}\ \bibnamefont
  {Tsuji}}, \bibinfo {author} {\bibfnamefont {Takashi}\ \bibnamefont {Oka}}, \
  and\ \bibinfo {author} {\bibfnamefont {Hideo}\ \bibnamefont {Aoki}},\
  }\bibfield  {title} {\enquote {\bibinfo {title} {Correlated electron systems
  periodically driven out of equilibrium: $\text{Floquet}+\text{DMFT}$
  formalism},}\ }\href {\doibase 10.1103/PhysRevB.78.235124} {\bibfield
  {journal} {\bibinfo  {journal} {Phys. Rev. B}\ }\textbf {\bibinfo {volume}
  {78}},\ \bibinfo {pages} {235124} (\bibinfo {year} {2008})}\BibitemShut
  {NoStop}%
\bibitem [{\citenamefont {Keller}(2011)}]{keller2011quantum}%
  \BibitemOpen
  \bibfield  {author} {\bibinfo {author} {\bibfnamefont {Ole}\ \bibnamefont
  {Keller}},\ }\href@noop {} {\emph {\bibinfo {title} {Quantum theory of
  near-field electrodynamics}}}\ (\bibinfo  {publisher} {Springer, Berlin},\
  \bibinfo {year} {2011})\BibitemShut {NoStop}%
\bibitem [{\citenamefont {Keldysh}(1965)}]{keldysh1965}%
  \BibitemOpen
  \bibfield  {author} {\bibinfo {author} {\bibfnamefont {L.~V.}\ \bibnamefont
  {Keldysh}},\ }\bibfield  {title} {\enquote {\bibinfo {title} {Diagram
  technique for nonequilibrium processes},}\ }\href
  {https://doi.org/10.1142/9789811279461_0007} {\bibfield  {journal} {\bibinfo
  {journal} {Soviet. Phys. JETP}\ }\textbf {\bibinfo {volume} {20}} (\bibinfo
  {year} {1965})}\BibitemShut {NoStop}%
\bibitem [{\citenamefont {Dyson}(1949)}]{dyson1949s}%
  \BibitemOpen
  \bibfield  {author} {\bibinfo {author} {\bibfnamefont {Freeman~J}\
  \bibnamefont {Dyson}},\ }\bibfield  {title} {\enquote {\bibinfo {title} {The
  {S} matrix in quantum electrodynamics},}\ }\href {\doibase
  10.1103/PhysRev.75.1736} {\bibfield  {journal} {\bibinfo  {journal} {Phys.
  Rev.}\ }\textbf {\bibinfo {volume} {75}},\ \bibinfo {pages} {1736} (\bibinfo
  {year} {1949})}\BibitemShut {NoStop}%
\bibitem [{\citenamefont {Meir}\ and\ \citenamefont
  {Wingreen}(1992)}]{meir1992landauer}%
  \BibitemOpen
  \bibfield  {author} {\bibinfo {author} {\bibfnamefont {Yigal}\ \bibnamefont
  {Meir}}\ and\ \bibinfo {author} {\bibfnamefont {Ned~S}\ \bibnamefont
  {Wingreen}},\ }\bibfield  {title} {\enquote {\bibinfo {title} {Landauer
  formula for the current through an interacting electron region},}\ }\href
  {\doibase 10.1103/PhysRevLett.68.2512} {\bibfield  {journal} {\bibinfo
  {journal} {Phys. Rev. Lett.}\ }\textbf {\bibinfo {volume} {68}},\ \bibinfo
  {pages} {2512} (\bibinfo {year} {1992})}\BibitemShut {NoStop}%
\bibitem [{\citenamefont {Kohler}\ \emph {et~al.}(2005)\citenamefont {Kohler},
  \citenamefont {Lehmann},\ and\ \citenamefont
  {H{\"a}nggi}}]{kohler2005driven}%
  \BibitemOpen
  \bibfield  {author} {\bibinfo {author} {\bibfnamefont {Sigmund}\ \bibnamefont
  {Kohler}}, \bibinfo {author} {\bibfnamefont {J{\"o}rg}\ \bibnamefont
  {Lehmann}}, \ and\ \bibinfo {author} {\bibfnamefont {Peter}\ \bibnamefont
  {H{\"a}nggi}},\ }\bibfield  {title} {\enquote {\bibinfo {title} {Driven
  quantum transport on the nanoscale},}\ }\href {\doibase
  10.1016/j.physrep.2004.11.002} {\bibfield  {journal} {\bibinfo  {journal}
  {Phys. Rep.}\ }\textbf {\bibinfo {volume} {406}},\ \bibinfo {pages}
  {379--443} (\bibinfo {year} {2005})}\BibitemShut {NoStop}%
\end{thebibliography}%


\begin{thebibliography}{10}%
\makeatletter
\providecommand \@ifxundefined [1]{%
 \@ifx{#1\undefined}
}%
\providecommand \@ifnum [1]{%
 \ifnum #1\expandafter \@firstoftwo
 \else \expandafter \@secondoftwo
 \fi
}%
\providecommand \@ifx [1]{%
 \ifx #1\expandafter \@firstoftwo
 \else \expandafter \@secondoftwo
 \fi
}%
\providecommand \natexlab [1]{#1}%
\providecommand \enquote  [1]{``#1''}%
\providecommand \bibnamefont  [1]{#1}%
\providecommand \bibfnamefont [1]{#1}%
\providecommand \citenamefont [1]{#1}%
\providecommand \href@noop [0]{\@secondoftwo}%
\providecommand \href [0]{\begingroup \@sanitize@url \@href}%
\providecommand \@href[1]{\@@startlink{#1}\@@href}%
\providecommand \@@href[1]{\endgroup#1\@@endlink}%
\providecommand \@sanitize@url [0]{\catcode `\\12\catcode `\$12\catcode
  `\&12\catcode `\#12\catcode `\^12\catcode `\_12\catcode `\%12\relax}%
\providecommand \@@startlink[1]{}%
\providecommand \@@endlink[0]{}%
\providecommand \url  [0]{\begingroup\@sanitize@url \@url }%
\providecommand \@url [1]{\endgroup\@href {#1}{\urlprefix }}%
\providecommand \urlprefix  [0]{URL }%
\providecommand \Eprint [0]{\href }%
\providecommand \doibase [0]{http://dx.doi.org/}%
\providecommand \selectlanguage [0]{\@gobble}%
\providecommand \bibinfo  [0]{\@secondoftwo}%
\providecommand \bibfield  [0]{\@secondoftwo}%
\providecommand \translation [1]{[#1]}%
\providecommand \BibitemOpen [0]{}%
\providecommand \bibitemStop [0]{}%
\providecommand \bibitemNoStop [0]{.\EOS\space}%
\providecommand \EOS [0]{\spacefactor3000\relax}%
\providecommand \BibitemShut  [1]{\csname bibitem#1\endcsname}%
\let\auto@bib@innerbib\@empty
\bibitem [{\citenamefont {Tsuji}\ \emph {et~al.}(2008)\citenamefont {Tsuji},
  \citenamefont {Oka},\ and\ \citenamefont {Aoki}}]{tsuji2008correlated}%
  \BibitemOpen
  \bibfield  {author} {\bibinfo {author} {\bibfnamefont {N.}~\bibnamefont
  {Tsuji}}, \bibinfo {author} {\bibfnamefont {T.}~\bibnamefont {Oka}}, \ and\
  \bibinfo {author} {\bibfnamefont {H.}~\bibnamefont {Aoki}},\ }\href {\doibase
  10.1103/PhysRevB.78.235124} {\bibfield  {journal} {\bibinfo  {journal} {Phys.
  Rev. B}\ }\textbf {\bibinfo {volume} {78}},\ \bibinfo {pages} {235124}
  (\bibinfo {year} {2008})}\BibitemShut {NoStop}%
\bibitem [{\citenamefont {Wang}\ \emph {et~al.}(2023)\citenamefont {Wang},
  \citenamefont {Peng}, \citenamefont {Zhang}, \citenamefont {Zhang},\ and\
  \citenamefont {Zhu}}]{wang2023transport}%
  \BibitemOpen
  \bibfield  {author} {\bibinfo {author} {\bibfnamefont {J.-S.}\ \bibnamefont
  {Wang}}, \bibinfo {author} {\bibfnamefont {J.}~\bibnamefont {Peng}}, \bibinfo
  {author} {\bibfnamefont {Z.-Q.}\ \bibnamefont {Zhang}}, \bibinfo {author}
  {\bibfnamefont {Y.-M.}\ \bibnamefont {Zhang}}, \ and\ \bibinfo {author}
  {\bibfnamefont {T.}~\bibnamefont {Zhu}},\ }\href {\doibase
  /10.1007/s11467-023-1260-z} {\bibfield  {journal} {\bibinfo  {journal}
  {Front. Phys.}\ }\textbf {\bibinfo {volume} {18}},\ \bibinfo {pages} {43602}
  (\bibinfo {year} {2023})}\BibitemShut {NoStop}%
\bibitem [{\citenamefont {Keller}(2011)}]{keller2011quantum}%
  \BibitemOpen
  \bibfield  {author} {\bibinfo {author} {\bibfnamefont {O.}~\bibnamefont
  {Keller}},\ }\href@noop {} {\emph {\bibinfo {title} {Quantum theory of
  near-field electrodynamics}}}\ (\bibinfo  {publisher} {Springer, Berlin},\
  \bibinfo {year} {2011})\BibitemShut {NoStop}%
\bibitem [{\citenamefont {Griffiths}(2017)}]{griffiths2023}%
  \BibitemOpen
  \bibfield  {author} {\bibinfo {author} {\bibfnamefont {D.~J.}\ \bibnamefont
  {Griffiths}},\ }\href@noop {} {\emph {\bibinfo {title} {Introduction to
  electrodynamics}}}\ (\bibinfo  {publisher} {Cambridge University Press},\
  \bibinfo {year} {2017})\BibitemShut {NoStop}%
\bibitem [{\citenamefont {Zhang}\ \emph {et~al.}(2022)\citenamefont {Zhang},
  \citenamefont {Zhu}, \citenamefont {Zhang},\ and\ \citenamefont
  {Wang}}]{zhang2022microscopic}%
  \BibitemOpen
  \bibfield  {author} {\bibinfo {author} {\bibfnamefont {Y.-M.}\ \bibnamefont
  {Zhang}}, \bibinfo {author} {\bibfnamefont {T.}~\bibnamefont {Zhu}}, \bibinfo
  {author} {\bibfnamefont {Z.-Q.}\ \bibnamefont {Zhang}}, \ and\ \bibinfo
  {author} {\bibfnamefont {J.-S.}\ \bibnamefont {Wang}},\ }\href {\doibase
  10.1103/PhysRevB.105.205421} {\bibfield  {journal} {\bibinfo  {journal}
  {Phys. Rev. B}\ }\textbf {\bibinfo {volume} {105}},\ \bibinfo {pages}
  {205421} (\bibinfo {year} {2022})}\BibitemShut {NoStop}%
\bibitem [{\citenamefont {Keldysh}(1965)}]{keldysh1965}%
  \BibitemOpen
  \bibfield  {author} {\bibinfo {author} {\bibfnamefont {L.~V.}\ \bibnamefont
  {Keldysh}},\ }\href {https://doi.org/10.1142/9789811279461_0007} {\bibfield
  {journal} {\bibinfo  {journal} {Soviet. Phys. JETP}\ }\textbf {\bibinfo
  {volume} {20}} (\bibinfo {year} {1965})}\BibitemShut {NoStop}%
\bibitem [{\citenamefont {Dyson}(1949)}]{dyson1949s}%
  \BibitemOpen
  \bibfield  {author} {\bibinfo {author} {\bibfnamefont {F.~J.}\ \bibnamefont
  {Dyson}},\ }\href {\doibase 10.1103/PhysRev.75.1736} {\bibfield  {journal}
  {\bibinfo  {journal} {Phys. Rev.}\ }\textbf {\bibinfo {volume} {75}},\
  \bibinfo {pages} {1736} (\bibinfo {year} {1949})}\BibitemShut {NoStop}%
\bibitem [{\citenamefont {Meir}\ and\ \citenamefont
  {Wingreen}(1992)}]{meir1992landauer}%
  \BibitemOpen
  \bibfield  {author} {\bibinfo {author} {\bibfnamefont {Y.}~\bibnamefont
  {Meir}}\ and\ \bibinfo {author} {\bibfnamefont {N.~S.}\ \bibnamefont
  {Wingreen}},\ }\href {\doibase 10.1103/PhysRevLett.68.2512} {\bibfield
  {journal} {\bibinfo  {journal} {Phys. Rev. Lett.}\ }\textbf {\bibinfo
  {volume} {68}},\ \bibinfo {pages} {2512} (\bibinfo {year}
  {1992})}\BibitemShut {NoStop}%
\bibitem [{\citenamefont {Kohler}\ \emph {et~al.}(2005)\citenamefont {Kohler},
  \citenamefont {Lehmann},\ and\ \citenamefont
  {H{\"a}nggi}}]{kohler2005driven}%
  \BibitemOpen
  \bibfield  {author} {\bibinfo {author} {\bibfnamefont {S.}~\bibnamefont
  {Kohler}}, \bibinfo {author} {\bibfnamefont {J.}~\bibnamefont {Lehmann}}, \
  and\ \bibinfo {author} {\bibfnamefont {P.}~\bibnamefont {H{\"a}nggi}},\
  }\href {\doibase 10.1016/j.physrep.2004.11.002} {\bibfield  {journal}
  {\bibinfo  {journal} {Phys. Rep.}\ }\textbf {\bibinfo {volume} {406}},\
  \bibinfo {pages} {379} (\bibinfo {year} {2005})}\BibitemShut {NoStop}%
\bibitem [{\citenamefont {Pan}\ \emph {et~al.}(2025)\citenamefont {Pan},
  \citenamefont {Ren}, \citenamefont {Tang},\ and\ \citenamefont
  {Wang}}]{Pan2025Asymmetry}%
  \BibitemOpen
  \bibfield  {author} {\bibinfo {author} {\bibfnamefont {H.}~\bibnamefont
  {Pan}}, \bibinfo {author} {\bibfnamefont {Y.}~\bibnamefont {Ren}}, \bibinfo
  {author} {\bibfnamefont {G.}~\bibnamefont {Tang}}, \ and\ \bibinfo {author}
  {\bibfnamefont {J.-S.}\ \bibnamefont {Wang}},\ }\href {\doibase
  10.1103/74rq-f642} {\bibfield  {journal} {\bibinfo  {journal} {Phys. Rev. B}\
  }\textbf {\bibinfo {volume} {112}},\ \bibinfo {pages} {L041401} (\bibinfo
  {year} {2025})}\BibitemShut {NoStop}%
\end{thebibliography}%

\end{document}


\title{Supplemental Material for ``Enhancing far-field thermal radiation by Floquet engineering"}
\author{Huimin Zhu,$^{1,2}$ Yuhua Ren,$^{2}$ Hui Pan,$^{2}$ Gaomin Tang,$^{3}$ Lei Zhang,$^{1,4}$ and Jian-Sheng Wang$^{2,*}$}
\affiliation{$^1$State Key Laboratory of Quantum Optics Technologies and Devices, Institute of Laser Spectroscopy, Shanxi University, Taiyuan 030006, China\\
	$^2$Department of Physics, National University of Singapore, Singapore 117551, Republic of Singapore\\
	$^3$Graduate School of China Academy of Engineering Physics, Beijing 100193, China\\
	$^4$Collaborative Innovation Center of Extreme Optics, Shanxi University, Taiyuan 030006, China}

\bigskip

\begin{abstract}
Section I provides a brief introduction of the Floquet representation. In Section II, we derive the free photon Green’s function, followed by the formulation of the retarded Green’s function in Section III. The expressions for photon energy currents are developed in Section IV. Finally, Section V presents the dispersion relations of surface phonon polaritons supported by the SiC film.
\end{abstract}

\maketitle

\section{I. Floquet representation}
For a time translationally invariant function $F(t, t')$ that depends only on the difference $t-t'$, it is customary to apply a Fourier transform and work in the frequency domain. However, if $F(t, t')$ possesses only discrete time-translation symmetry, satisfying $F(t+2\pi/\Omega, t'+2\pi/\Omega) = F(t, t')$, it can still be analyzed within the Floquet framework~\cite{tsuji2008correlated}, as defined by
\begin{equation}
    F_{mn}(\omega) = \frac{\Omega}{2\pi} \int_0^{2\pi/\Omega} d t \int_{-\infty}^\infty d t' e^{i (\omega_m t - \omega_n t')} F(t, t') ,
\end{equation}
where we define the shorthand $\omega_m = \omega + m \Omega$. Each $F_{mn}(\omega)$ can be interpreted as the element of a block matrix $\bm{F}(\omega)$. The corresponding Fourier inversion is
\begin{equation}
    F(t, t') = \sum_{m,n \in \mathbb{Z}} \int_{-\Omega/2}^{\Omega/2} \frac{d\omega}{2\pi} F_{mn}(\omega) e^{-i (\omega_m t - \omega_n t')} .
\end{equation}
For $t=t'$, the time average of $F(t, t)$ with period $\tau = 2\pi / \Omega$ can be obtained as
\begin{align}\label{Ftt}
{F} \equiv  \frac{1}{\tau } \int_{0}^\tau  F(t,t) dt  = \sum_n \int_{-\Omega/2}^{\Omega/2} \frac{d\omega}{2\pi}F_{nn}(\omega).
\end{align}
The Floquet representation allows us to preserve many useful properties of the usual Fourier transform. For example, differentiation is multiplication by a block diagonal matrix in Floquet representation,
\begin{align}
    i\hbar \frac{\partial}{\partial t} F(t, t') &\to \bm{\mathcal{E}}(\omega) \bm{F}(\omega),  \label{FE} \\
    i\hbar \frac{\partial}{\partial t'} F(t, t') &\to - \bm{F}(\omega) \bm{\mathcal{E}}(\omega),
\end{align}
where the components are $\mathcal{E}_{mn}(\omega) = \delta_{mn} \hbar \omega_m$. Another useful application of the Floquet representation is the convolution theorem~\cite{tsuji2008correlated}, where convolution in time becomes block matrix multiplication in Floquet representation,
\begin{align}\label{Ctt}
    C(t, t') = \int_{-\infty}^\infty d t'' A(t, t'') B(t'', t') \to \bm{C}(\omega) = \bm{A}(\omega) \bm{B}(\omega).
\end{align}

\section{II. Free photon Green's function}
In the temporal gauge, the current can be expressed in terms of the vector potential due to Maxwell's equations,
\begin{equation}
    \mathbf{J} = \epsilon_0 \biggl[ c^2 \nabla \times (\nabla \times \mathbf{A}) + \frac{\partial^2}{\partial t^2} \mathbf{A} \biggr] = - v^{-1} \mathbf{A} . \label{eqn:maxwell4}
\end{equation}
The differential operator $v^{-1}$ is defined with the negative sign consistent with the usual nonequilibrium Green's function  convention~\cite{wang2023transport}. The free photon Green's function is the solution to the same differential equation with a delta source,
\begin{equation}
    v^{-1} v^r(\mathbf{r}, t) = \delta^{(3)}(\mathbf{r}) \delta(t) I .
\end{equation}
For systems with translational symmetry in space and time, such as a homogeneous medium in thermal equilibrium, it is more convenient to perform the Fourier transform, where the expression can be more easily inverted. Due to the planar geometry of the SiC film, we can only carry out the spatial Fourier transform in the in-plane directions to replace the position variables $(x,y)$ with the corresponding in-plane wavevector ${\mathbf q}_\perp = (q_x, q_y)$.
\begin{equation}
    v^r(\mathbf{q}_\perp, \omega, z) = \int dt dx dy \, e^{i (\omega t - \mathbf{q}_\perp \cdot \mathbf{r})} v^r(\mathbf{r}, t).
\end{equation}
The operator $v^{-1}$ after Fourier transform becomes
\begin{equation} \label{v}
    v^{-1} (\mathbf{q}_\perp, \omega, z)
    = -\epsilon_0 \left[ c^2 \begin{pmatrix}
        q_y^2 - \frac{\partial^2}{\partial z^2} & -q_x q_y & iq_x \frac{\partial}{\partial z} \\
        -q_x q_y & q_x^2 - \frac{\partial^2}{\partial z^2} & iq_y \frac {\partial}{\partial z} \\
        iq_x \frac{\partial}{\partial z} & iq_y \frac {\partial}{\partial z} & q_x^2 + q_y^2
    \end{pmatrix} - \omega^2 \begin{pmatrix}
        1 & 0 & 0 \\
        0 & 1 & 0 \\
        0 & 0 & 1
    \end{pmatrix} \right].
\end{equation}
In this mixed representation, the (retarded) Green's function is given by~\cite{wang2023transport,keller2011quantum}
\begin{equation}
    v^r(\mathbf{q}_\perp, \omega, z) = \frac{\mu_0}{k_0^2}
    \left\{ \frac{e^{i k |z|}}{2i k}
    \begin{bmatrix}
        k_0^2 - q_x^2 & -q_x q_y & - \text{sgn}(z) q_x k \\
        -q_x q_y & k_0^2 - q_y^2 & - \text{sgn}(z) q_y k \\
        -\text{sgn}(z) q_x k & - \text{sgn}(z) q_y k & q_\perp^2
    \end{bmatrix}
    + \delta(z) \hat{\mathbf{z}} \hat{\mathbf{z}} \right\} .
\end{equation}
Here, $k = \sqrt{k_0^2 - q_\perp^2}$ is the propagation constant in the $z$ direction, with $k_0 = \omega/c$ and $q_\perp = |\mathbf{q}_\perp|$. In the evanescent case where $k_0 < q_\perp$, the sign of the square root is taken so that the imaginary part is positive, i.e. $k = i \sqrt{q_\perp^2 - k_0^2}$.

\section{III. Solving the Dyson equation in Floquet representation}
\label{sec:Dyson}
A key piece of information in the Dyson equation is the self-energy of the object under periodic modulation.
The electric susceptibility of the SiC film $\chi(t,t') = \chi_0(t-t') + \chi_d(t,t')$, is given by the sum of the time-invariant component, which is related to equilibrium permittivity by $\chi_0(t-t') = \epsilon(t-t') - 1$,  and the time-modulated component $\chi_d(t,t') = 2 \Delta \chi \delta(t-t') \cos(\Omega t)$. In the Floquet matrix representation, they can be respectively expressed as diagonal and tridiagonal,
\begin{equation}\label{chi}
\bm{\chi}_0(\omega) =
\begin{pmatrix}
\ddots & & & & \\
 & \epsilon(\omega-\Omega)-1 & & & \\
 &  & \epsilon(\omega)-1 & & \\
 &  &  & \epsilon(\omega+\Omega)-1 & \\
 &  &  &  & \ddots
\end{pmatrix},
\quad\quad
\bm{\chi}_d(\omega) =
\begin{pmatrix}
\ddots & \ddots & & & \\
 \ddots & 0 & \Delta\chi & & \\
  & \Delta\chi & 0 & \Delta\chi & \\
 &  & \Delta\chi & 0 &   \ddots\\
 &  &  & \ddots & \ddots
\end{pmatrix},
\end{equation}
so that the total susceptibility in Floquet representation is $\bm{\chi}(\omega) = \bm{\chi}_0(\omega) + \bm{\chi}_d(\omega)$. It is assumed that the SiC film is isotropic in the $xy$ plane, so that the permittivity can be treated as a scalar quantity.

The self-energy can be defined from the perspective of linear response, as the Green's function for the induced polarization current due to a vector potential,
\begin{equation}
    \frac{\partial}{\partial t}\biggl( \int d t' \epsilon_0 \chi(t, t') \mathbf{E}(t') \biggr) = \frac{\partial}{\partial t}{\mathbf{P}(t)} = - \frac{1}{a} \int d t' \Pi^r(t, t') \mathbf{A}(t') ,
\end{equation}
where $a$ is the thickness of the film, $\epsilon_0$ is the vacuum permittivity, and $\mathbf{P}$ is the polarization vector (dipole moment per unit volume). Using integration by parts, the self-energy is then identified as
\begin{equation}\label{PIrt}
    \Pi^r(t, t') = -a \epsilon_0 \frac{\partial}{\partial t} \frac{\partial}{\partial t'} \chi(t, t') .
\end{equation}
Thus, the self-energy functions can be written in Floquet representation as
\begin{align}
\mathbf{\Pi}^r_0 &= - a \epsilon_0 \bm{\mathcal{E}} \bm{\chi}_0 \bm{\mathcal{E}}/\hbar^2, \label{PIr0} \\
\mathbf{\Pi}^r_d &= - a \epsilon_0 \bm{\mathcal{E}} \bm{\chi}_d \bm{\mathcal{E}}/\hbar^2, \label{PIrd} \\
\mathbf{\Pi}^r &= - a \epsilon_0 \bm{\mathcal{E}} \bm{\chi} \bm{\mathcal{E}}/\hbar^2.\label{PIr}
\end{align}
With the self-energy functions, we can write down the Dyson equation, which in differential form is
\begin{align}
v^{-1}D^r(\mathbf{r},t, \mathbf{r}',t') = I\delta(\mathbf{r}-\mathbf{r}')\delta(t-t') + \int d\mathbf{r}'' \int dt'' \Pi^r\left(\mathbf{r},t, \mathbf{r}'', t'' \right) D^r \left(\mathbf{r}'',t'', \mathbf{r}',t' \right).
\end{align}
We perform a spatial Fourier transform in the $xy$ direction to obtain
\begin{align}
    v^{-1}D^r(\mathbf{q}_\perp, z,t, z',t') = I\delta(z-z')\delta(t-t') + \int dz'' \int dt'' \Pi^r \left(\mathbf{q}_\perp, z,t, z'',t'' \right) D^r \left(\mathbf{q}_\perp, z'',t'', z',t' \right).
\end{align}
The SiC film is located at $z = 0$, so the self-energy takes the form of $\Pi^r \left(\mathbf{q}_\perp, z,t, z'',t'' \right) = \Pi^r \left(\mathbf{q}_\perp,t, t'' \right)\delta(z)\delta(z'')$. Expressing the Dyson equation in Floquet representation and using the convolution theorem, we have
\begin{align}\label{Dr}
v^{-1}\bm{D}^r(\mathbf{q}_\perp, \omega, z, z') = \bm{I} \delta(z-z') + \delta(z) \bm{\Pi}^r\left(\mathbf{q}_\perp, \omega \right) \bm{D}^r \left(\mathbf{q}_\perp,\omega, 0, z' \right).
\end{align}
Substituting Eq.~\eqref{v} into Eq.~\eqref{Dr}, and considering that the material is isotropic in the $xy$ plane, we solve the components of the photon Green’s function
for $s$-polarization and $p$-polarization, respectively, by rotating the Cartesian coordinate system, and finally obtain
\begin{align}
    \bm{D}^r(\mathbf{q}_\perp, \omega, z, 0)
    =& \frac{1}{2i \epsilon_0 c^2 q_\perp^2}
    \begin{pmatrix}
        q_y^2 & -q_x q_y & 0 \\
        -q_x q_y & q_x^2 & 0 \\
        0 & 0 & 0
    \end{pmatrix}
    e^{i{\bm k}|z|} \bm{k}^{-1} {\bm t^s}  \notag \\
    & + \frac{\hbar^2}{2i \epsilon_0 q_\perp^2 \bm{\mathcal{E}}^2}
    \begin{pmatrix}
        q_x^2 \bm{k}^2 & q_x q_y \bm{k}^2 & -q_x q_\perp^2 \bm{k} \\
        q_x q_y \bm{k}^2 & q_y^2 \bm{k}^2& -q_y q_\perp^2 \bm{k} \\
        -q_x q_\perp^2 \bm{k} & -q_y q_\perp^2 \bm{k} & q_\perp^4
    \end{pmatrix}
    e^{i{\bm k}|z|}\bm{k}^{-1} {\bm t^p},
\end{align}
with the transmission coefficients for $s$ and $p$ polarizations being
\begin{align}
    \bm{t}^s &= \big[ \mathbf{I} - {i a}/{(2 \hbar^2 c^2)} \bm{\mathcal{E}} \bm{\chi} \bm{\mathcal{E}} \bm{k}^{-1} \big]^{-1}, \\
    \bm{t}^p &= \big[ \mathbf{I} - {i a} \bm{\mathcal{E}} \bm{\chi} \bm{\mathcal{E}}^{-1} \bm{k}/{2} \big]^{-1}.
\end{align}
Here, $\bm{\chi}$ is the total susceptibility in Floquet representation, and $\bm{k}$ is the Floquet generalization of $k_0$, with elements $k_{mn}=\delta_{mn} k_n$, where
\begin{equation}
    k_n = \begin{cases}
        \text{sgn}(\omega_n)\sqrt{(\omega_n/c)^2-q_\perp^2},
        &\text{if } |\omega_n| > cq_\perp, \\
        \sqrt{(\omega_n/c)^2-q_\perp^2},
        &\text{if } |\omega_n| < cq_\perp.
    \end{cases}
\end{equation}

\section{IV. Derivation of the expressions of photon energy currents}
To obtain the energy transfer of the system, we start from Poynting's theorem~\cite{griffiths2023},  which in differential form is
\begin{align}
\frac{\partial u}{\partial t} + \nabla \cdot \mathbf{S} = -\mathbf{J} \cdot \mathbf{E}.
\end{align}
Here, $u = (\epsilon_0 E^2 + B^2 /\mu_0) /2$ is the field energy density, and $\mathbf{S}= \mathbf{E} \times \mathbf{B} /\mu_0$ is the Poynting vector. In the temporal gauge, $\mathbf{E}=-{\partial \mathbf{A}} /{\partial t}$, $\mathbf{B}= \nabla \times \mathbf{A}$ with the vector potential $\mathbf{A}$. The electric current density $\mathbf{J}$ can be expressed in terms of the vector potential as~\cite{zhang2022microscopic} $\mathbf{J} = - v^{-1} \mathbf{A}$ with the differential operator $v^{-1} = -\epsilon_0 \partial^2 /{\partial t^2} - \mu_0^{-1} \nabla \times (\nabla \times)$, where $\epsilon_0$ is the vacuum permittivity and $\mu_0$ is the vacuum permeability.

In a steady-state system, the time average of the energy density vanishes, i.e., $\langle {\partial u}/{\partial t} \rangle = 0$. The net energy flux through the enclosed surface is entirely contributed by Joule heating. Using the divergence theorem, the surface integrals can be transformed into volume integrals over the object $\alpha$, and the thermal energy emitted by the object is expressed as
\begin{align}
I_\alpha(t) \equiv \oint_\alpha \langle \mathbf{S} \rangle \cdot d\mathbf{\Sigma}
= - \int_\alpha d\mathbf{r} \langle \mathbf{E} \cdot \mathbf{J} \rangle.
\end{align}
The index $\alpha$ refers to either the SiC film or the baths at infinity. $\langle\cdots\rangle$ denotes the statistical ensemble average, and $d\mathbf{\Sigma}$ is the surface element with an outward normal.

For a periodically driven system, the thermal energy emission $I_\alpha(t)$ can be expressed in terms of the Keldysh Green's function as
\begin{align}\label{Ia-1}
I_\alpha(t) =& - \mathlarger{\int}_\alpha d\mathbf{r} \biggl[ \frac{\partial}{\partial t} \sum_{\mu,\nu}  \bigl\langle A_\mu(\mathbf{r},t) A_\nu(\mathbf{r}',t') \bigr\rangle v^{-1}_{\nu \mu}(\mathbf{r}',t')  \biggr] \biggr|_{t'=t,\mathbf{r'}=\mathbf{r}}
\notag \\
=& - \frac{i\hbar}{2} \mathlarger{\int}_\alpha d\mathbf{r}
\biggl\{ \frac{\partial}{\partial t} {\text {tr}}  \bigl[  D^K(\mathbf{r},t,\mathbf{r}',t')  v^{-1}(\mathbf{r}',t') \bigr] \biggr\}
\biggr|_{t'=t,\mathbf{r'}=\mathbf{r}},
\end{align}
where $\mu$, $\nu$ $\in$ \{$x$, $y$, $z$\}, and the trace is over directions. The Keldysh Green's function is $D^K = D^< + D^>$. The lesser and greater Green's functions are given by the Keldysh equation~\cite{keldysh1965} $D^{<(>)} = D^r \Pi^{<(>)} D^a$. Here, $\Pi^{<(>)}$ is the self-energy, and $D^r(D^a)$ is the retarded (advanced) Green's function. In the frequency domain, we have  $({D}^r)^{\dagger}={D}^a$. This general relation is also shared by the self-energy. The retarded Green's function satisfies the Dyson equation~\cite{dyson1949s} $v^{-1} D^r = I + \Pi^r D^r$, where $\Pi^r$ is the retarded self-energy. For the time-modulated SiC film, its retarded self-energy function is given by $\Pi^r(t, t') = -a \epsilon_0 ({\partial}^2/{\partial t} {\partial t'}) [\chi_0(t- t')+\chi_d(t, t')]$ [see Eq.~\eqref{PIrt}], indicating that the self-energy consists of contributions from both the equilibrium and the time-modulated parts.

Using the Dyson equation $v^{-1} D^a = I + \Pi^a D^a$, and the Keldysh equation $D^K = D^r \Pi^K D^a$, we find $D^Kv^{-1} = D^r \Pi^K +D^K\Pi^a$. By substituting this into Eq.~\eqref{Ia-1}, we obtain the Meir-Wingreen formula~\cite{meir1992landauer} for energy current as
\begin{align}
I_{\alpha}(t)=-\frac{i\hbar}{2} \int_{\alpha} d\mathbf{r} \int d\mathbf{r}_1 \int dt_1 \frac{\partial}{\partial t} \text{tr} \Big[ D^{r}(\mathbf{r},t, \mathbf{r}_1,t_1)\Pi^{K}_{\alpha}(\mathbf{r}_1,t_1,\mathbf{r}',t')
+D^{K}(\mathbf{r},t, \mathbf{r}_1,t_1) \Pi^{a}_{\alpha}(\mathbf{r}_1,t_1,\mathbf{r}',t') \Big] \Big|_{t'=t,\mathbf{r'}=\mathbf{r}}.
\end{align}
Expressing the energy current in Floquet representation and using Eqs.~\eqref{Ftt}, \eqref{FE}, and \eqref{Ctt}, the average energy current ${I}_{\alpha}$ in the frequency domain can be expressed as
\begin{align}\label{Ia-2}
{I}_\alpha = - \int_{-\Omega/2}^{\Omega/2} \frac{d\omega}{4\pi}
{\rm Tr} \bigg[ \bm{\mathcal{E}} \Big( \bm{D}^r \bm{\Pi}^K_\alpha + \bm{D}^K \bm{\Pi}^a_\alpha \Big) \bigg],
\end{align}
where the symbol ``Tr'' denotes tracing over the directions, positions, and Floquet indices. The superscripts $K$, $r$, and $a$ denote the Keldysh, retarded, and advanced components, respectively. We use the boldface letters to denote the matrices in Floquet space. The entries of the matrix $\bm{\mathcal{E}}$ are $\mathcal{E}_n\delta_{mn}$, where $\mathcal{E}_n=\hbar\omega+n\hbar\Omega$ with $n \in \mathbb{Z}$. The $\bm{D}$ and $\mathbf{\Pi}$ are Floquet matrices of the photon Green's function and self-energy, respectively. The expression for $\bm{D}^r$ in our system is given in Section III.

For a system with periodic time modulation, we assume local thermal equilibrium for both the equilibrium part, denoted as $\mathcal{O}$ and the baths at infinity, denoted as $\infty$, which allows the application of the fluctuation-dissipation theorem to the self-energies. In the nonequilibrium Green's function formalism, this is given by $\Pi_{\alpha}^K = (2N_{\alpha}+1) \left( \Pi_{\alpha}^r - \Pi_{\alpha}^a \right)$, in a concise form for the materials properties. Here, $N_{\alpha} \equiv N(T_{\alpha},\omega) = \left[\exp(\hbar \omega/{k_B T_{\alpha}})-1 \right]^{-1}$ is the Bose-Einstein distribution function. However, the local thermal equilibrium assumption fails for the driven part due to ${\bm \chi}_d$ being dissipationless.

The energy current ${I}_\alpha$ is a physical observable, it must be real. Therefore, we can express it as the average of itself and its Hermitian conjugate,
\begin{align}\label{Ia-3}
{I}_\alpha = \frac{1}{2} \left[ {I}_\alpha + ({I}_\alpha)^{\dagger} \right].
\end{align}
We first focus on the energy currents for the equilibrium part and the baths at infinity. Equation~\eqref{Ia-3} can be simplified using the fact that (1) the symmetries of the Floquet representation $(\bm{D}^r)^\dagger=\bm{D}^a$ and $(\bm{D}^K)^\dagger=-\bm{D}^K$; (2) the relation $\bm{D}^r-\bm{D}^a=\bm{D}^>-\bm{D}^<$; (3) the commutative relation $[\bm{\mathcal{E}},\bm{\Pi}_{\alpha}^K]=0$;  (4) the fluctuation-dissipation theorem for the Keldysh self-energy function $\bm{\Pi}_{\alpha}^K= (2 {\bm N}_{\alpha} +1) \left( \bm{\Pi}_{\alpha}^r - \bm{\Pi}_{\alpha}^a \right)$, where $\bm N$ is the Bose-Einstein distribution function in matrix form; (5) we can perform cyclic permutation under trace; Furthermore, the relations (1) and (2) also apply to the self-energy $\bm{\Pi}$. With these relations and the manipulation, a Landauer formula~\cite{kohler2005driven} for the energy current can be obtained as
\begin{align}\label{Ia-4}
{I}_\alpha = \int_{-\Omega/2}^{\Omega/2} \frac{d\omega}{4\pi}
{\rm Tr} \sum\limits_{\beta}  \bigg[\bm{\mathcal{E}} \Big( \bm{D}^r \mathbf {\Gamma}_\beta \bm{D}^a \bm{N}_\alpha \mathbf{\Gamma}_\alpha
-\bm{D}^r \mathbf{\Gamma}_\beta \bm{N}_\beta \bm{D}^a \mathbf{\Gamma}_\alpha \Big) \bigg].
\end{align}
Here, we define $\mathbf{\Gamma}_{\alpha,\beta}=i\left(\mathbf{\Pi}^r_{\alpha,\beta}-\mathbf{\Pi}^a_{\alpha,\beta}\right)$. The indices $\alpha$ and $\beta$ refer to either the equilibrium part or the baths at infinity.  The self-energy of the equilibrium part is given by Eq.~\eqref{PIr0}, and the self-energy of the baths at infinity is given by
\begin{align}
{\bm \Pi}^r_\infty = -i \epsilon_0 c^2 \bm{k} \left[ {\mathbf I}_\perp + \frac{1}{ {\bm k}^2} \left( {\mathbf q}_\perp {\mathbf q}^{T}_\perp \right) \right].
\end{align}
Using the relation~\cite{Pan2025Asymmetry} $F_{mn}(\omega + k\Omega) = F_{m+k,n+k}(\omega)$, the frequency integration range of Eq.~\eqref{Ia-4} can be extended to the entire frequency domain, and the energy current ${I}_\alpha$ can be written as
\begin{align}
{I}_\alpha = \int_{-\infty}^{\infty} \frac{d\omega}{4\pi}
{\rm Tr} \sum\limits_{\beta}  \bigg[\hbar\omega \Big( \bm{D}^r \mathbf {\Gamma}_\beta \bm{D}^a \bm{N}_\alpha \mathbf{\Gamma}_\alpha
-\bm{D}^r \mathbf{\Gamma}_\beta \bm{N}_\beta \bm{D}^a \mathbf{\Gamma}_\alpha \Big)_{00} \bigg].
\end{align}
Here, the trace is taken over the ``00'' block of the Floquet matrix. Due to the symmetry of the integrand with respect to positive and negative frequencies, the integration range can be changed from ($-\infty$,$\infty$) to [0,$\infty$) and multiply by a factor of 2. Thus, we have the Landauer formula,
\begin{align} \label{Ia-6}
{I}_\alpha = \int_{0}^{\infty} \frac{d\omega}{2\pi} \hbar\omega \sum\limits_{\beta,n} {\mathcal T}^{\beta, \alpha}_{n} \big[  {N}(T_\alpha,\omega)- {N}(T_\beta,\omega_n) \big],
\end{align}
with 
\begin{align}\label{Ia-6}
{\mathcal T}^{\beta, \alpha}_{n} = {\rm tr} \big[ (\bm{D}^r)_{0n} (\mathbf{\Gamma}_\beta)_{nn} (\bm{D}^a)_{n0} (\mathbf{\Gamma}_\alpha)_{00} \big],
\end{align}
where the superscripts of ${\mathcal T}^{\beta, \alpha}$ indicate the transmission from subsystem $\alpha$ to subsystem $\beta$, while the subscripts ($0n$, $nn$, $n0$, $00$) on the right-hand side denote the block elements of the Floquet matrices.

The energy emitted by the bath at $+\infty$ can be obtained by
\begin{align}\label{Ia-7}
{I}_{+\infty} =& \int_{0}^{\infty} \frac{d\omega}{2\pi} \hbar\omega \sum\limits_{n} \Big\{ {\mathcal T}^{-\infty, +\infty}_{n} \big[ {N}(T_{ +\infty},\omega) - {N}(T_{-\infty}, \omega_n) \big]  \notag\\
  &+ {\mathcal T}^{+\infty,+\infty}_{n} \big[ {N}(T_{+\infty}, \omega) - {N}(T_{+\infty}, \omega_n) \big] + {\mathcal T}^{\mathcal{O},+\infty}_{n} \big[ {N}(T_{+\infty}, \omega) - {N}(T_{\mathcal{O}}, \omega_n) \big] \Big\},
\end{align}
where $N(T, \omega)$ is the Bose-Einstein distribution function at temperature $T$. The baths at infinity are at zero temperature, so that ${N}(T_{+\infty}=0, \omega) = 0$ with $\omega > 0$, and ${N}(T_{+\infty}=0, \omega_n) = 0$ if $\omega_n > 0$, while ${N}(T_{+\infty}=0, \omega_n) = -1$ if $\omega_n < 0$. Equation~\eqref{Ia-7} can be simplified to
\begin{align}\label{Ia-8}
{I}_{+\infty} = &  \int_{0}^{\infty} \frac{d\omega}{2\pi} \hbar\omega \sum\limits_{n} \Big[ {\mathcal T}^{-\infty, +\infty}_{n} \theta(-\omega_n) + {\mathcal T}^{+\infty,+\infty}_{n} \theta(-\omega_n) - {\mathcal T}^{\mathcal{O},+\infty}_{n} {N}(T_{\mathcal{O}}, \omega_n) \Big],
\end{align}
where $\theta(\omega)$ is the Heaviside step function. By inserting the photon Green's function ${\bm D}$ and the self-energy $\bm{\Pi}$ into the expression of ${\mathcal T}$ in Eq.~\eqref{Ia-8}, we can obtain the semi-analytical form for the energy emitted by the bath at $+\infty$ as
\begin{align}
{I}_{+\infty}=& \int_{0}^{\infty} \frac{d\omega}{2\pi} \hbar\omega \sum\limits_{n} \int_{|\mathbf{q}_\perp|<k_0} \frac{d^2 \mathbf{q}_\perp}{(2\pi)^2} \bigg[ \frac{ {\rm Im} [\Pi_0^r(\omega_n)]}{\epsilon_0 c^2 k} \left(|t^s_n|^2 + \frac{k^2 }{k_0^2} |t^p_n|^2\right)  N(T_{\mathcal{O}},\omega_n) - \left(\frac{k_n } { k } |t^s_n|^2 +\frac{k \omega_n^2}{k_n \omega^2}|t^p_n|^2  \right)\theta(-\omega_n)\bigg].
\end{align}
Given the spatial symmetry of the SiC film in the $z$-direction, the energy emitted by the bath at $+\infty$ is equal to the energy emitted by the bath at $-\infty$. Thus, the total energy emitted by the baths at infinity is ${I}_{\infty} = 2 {I}_{+\infty}$.

Similar to the calculation of ${I}_{+\infty}$, we calculate the energy emitted by the equilibrium part,
\begin{align}\label{Ia-9}
{I}_{\mathcal{O}} =& \int_{0}^{\infty} \frac{d\omega}{2\pi} \hbar\omega \sum\limits_{n} \Big\{{\mathcal T}^{-\infty, \mathcal{O}}_{n} \big[ {N}(T_{\mathcal{O}}, \omega) - {N}(T_{-\infty}, \omega_n) \big] \notag \\
  &+ {\mathcal T}^{+\infty, \mathcal{O}}_{n} \big[ {N}(T_{\mathcal{O}}, \omega) - {N}(T_{+\infty}, \omega_n) \big]
  + {\mathcal T}^{\mathcal{O}, \mathcal{O}}_{n} \big[ {N}(T_{\mathcal{O}},\omega) - {N}(T_{\mathcal{O}}, \omega_n) \big] \Big\} \notag \\
  =& \int_{0}^{\infty} \frac{d\omega}{2\pi} \frac{\hbar\omega}{\epsilon_0 c^2} \sum\limits_{n} \int_{|\mathbf{q}_\perp|<k_0} \frac{d^2 \mathbf{q}_\perp}{(2\pi)^2}  {\rm Im}[\Pi_0^r(\omega)] \Bigg\{
 \frac{ {\rm Im}[\Pi_0^r(\omega_n)]}{\epsilon_0c^2k^2}    \left(|t^s_n|^2+  \frac{k^4}{k_0^4}|t^p_n|^2  \right) \Big[ N(T_{\mathcal{O}},\omega)-N(T_{\mathcal{O}},\omega_n)\Big]   \notag \\
 & - \left( \frac{2k_n}{k^2} |t^s_n|^2+ \frac{2\omega_n^2 k^2} {\omega^2 k_0^2 k_n  } |t^p_n|^2 \right) \Big[ N(T_{\mathcal{O}},\omega)+\theta(-\omega_n)\Big]   \Bigg\}.
\end{align}
However, the fluctuation-dissipation theorem fails for the driven part, the energy pumped into the electromagnetic field due to time modulation needs to be returned to Eq.~\eqref{Ia-2} for further derivation. Combining with Eq.~\eqref{Ia-3}, we have
\begin{align}
{I}_d =& \frac{1}{2} \left[ {I}_d + ({I}_d)^{\dagger} \right] \notag \\
=&-\frac{1}{2} \int_{-\Omega/2}^{\Omega/2} \frac{d\omega}{4\pi}
{\rm Tr}  \Bigg\{ \bm{\mathcal{E}} \Big( \bm{D}^r \bm{\Pi}^K_d + \bm{D}^K \bm{\Pi}^a_d \Big)  + \bigg[ \bm{\mathcal{E}} \Big( \bm{D}^r \bm{\Pi}^K_d + \bm{D}^K \bm{\Pi}^a_d \Big) \bigg]^{\dagger} \Bigg\} \notag \\
=&  -\int_{-\Omega/2}^{\Omega/2}\frac{d\omega}{8\pi} {\rm Tr} \Bigg\{ [\mathbf{\Pi}^r_d,\bm{\mathcal{E}}] \sum\limits_{\beta \neq d} \bm{D}^r \left( 2\bm{N}_{\beta}+1 \right)  \left(\mathbf{\Pi}^r_{\beta}-\mathbf{\Pi}^a_{\beta}\right) \bm{D}^a \Bigg\} .
\end{align}
Here, we have used $\bm{\Pi}^r_d= \bm{\Pi}^{a}_d$ and  $\bm{\Pi}^K_d=0$.

\section{V. Dispersion relations of surface phonon polaritons supported by the SiC film}
\label{sec:sphp}
The dispersion relation of surface phonon polaritons (SPhPs) in a SiC film can be obtained by setting the dielectric function to zero,
\begin{equation}
    \begin{bmatrix}
        1 & 0 & 0  \\
        0 & 1 & 0  \\
        0 & 0 & 0
    \end{bmatrix} - v^r \Pi_0^r = 0.
\end{equation}
When $z=0$, the free photon Green's function satisfies
\begin{equation}
    v^r(\mathbf{q}_\perp, \omega, 0) = \frac{\mu_0}{2 i k k_0^2}
    \begin{bmatrix}
        k_0^2 - q_x^2 & -q_x q_y & 0  \\
        -q_x q_y & k_0^2 - q_y^2 & 0  \\
        0 & 0 & q_\perp^2
    \end{bmatrix} ,
\end{equation}
and the retarded polarization function is
\begin{equation}
\Pi_0^r(\omega)= -a\epsilon_0 \omega^2 \big[\epsilon(\omega) - 1 \big].
\end{equation}
For the equation to have a solution, either $q_x=0$ or $q_y = 0$. Without loss of generality, let $q_x = 0$, which means that the $s$-polarization is in the $x$ direction. We can obtain the dispersion relations for both the transverse electric ($s$-polarized) mode and transverse magnetic ($p$-polarized) mode.
\begin{equation}
\begin{bmatrix}
1+ \dfrac { a k_0^2 \big[\epsilon(\omega) -1 \big] } { 2 i k } & 0   \\
0 & 1 + \dfrac { a k \big[\epsilon(\omega) -1 \big] } { 2 i }   \\
\end{bmatrix}
=0.
\end{equation}
Therefore, the dispersion relations of the $s$- and $p$-polarized modes are as follows
\begin{equation}
\begin{aligned}
    q_\perp^2 &= k_0^2 + \frac{1}{4} a^2 \big[\epsilon(\omega) - 1 \big]^2 k_0^4 \quad \text{for $s$-polarized},  \qquad
    q_\perp^2 &= k_0^2 + \frac{4}{a^2 \big[ \epsilon(\omega) - 1 \big]^2} \quad \text{for $p$-polarized}.
\end{aligned}
\end{equation}
Note that the surface wave decays exponentially in the $z$ direction, and its dispersion relation must satisfy $ik<0$ , where $k=\sqrt{k_0^2-q_{\perp}^2}$. In particular, the $s$-polarized mode must satisfy $-ak_0^2 \big[\epsilon(\omega)-1 \big]/2<0$, while the $p$-polarized mode must satisfy $2/ \big[ a \left( \epsilon(\omega)-1 \right) \big]<0$.

\bibliography{SM_Farfield_Floquet}{}